\documentclass[aps,preprintnumbers,amsmath,amssymb,superscriptaddress,floatfix,nofootinbib]{revtex4}
\usepackage{cancel}
\usepackage{slashed}
\usepackage{graphicx}
\usepackage{epsfig} 
\usepackage{epstopdf}
\usepackage{hyperref}
\usepackage{amsfonts,color}
\usepackage{natbib}
\usepackage{subfig}
\usepackage{bbold}
\usepackage{bm}
\usepackage{mathrsfs}
\usepackage{url}
\usepackage{rotating}

\usepackage{multirow}
\usepackage{hhline}

\newcommand{\LO}{\ensuremath{\text{LO}}}
\newcommand{\NLO}{\ensuremath{\text{NLO}}}

\usepackage{pgfplots}

\begin{document}

\preprint{UWThPh 2025-21}

\title{On the existence of bound states in SIMP dark sectors}

\author{Xiaoyong Chu}
\email{chuxiaoyong@ucas.ac.cn}
\affiliation{International Centre for Theoretical Physics Asia-Pacific (ICTP-AP), University of Chinese Academy
of Sciences, 100190 Beijing, China}

\author{Josef Pradler}
\email{josef.pradler@univie.ac.at}
\affiliation{
Marietta Blau Institute for Particle Physics, Austrian Academy of Sciences,\\ Dominikanerbastei 16, A-1010 Vienna, Austria}
\affiliation{University of Vienna, Faculty of Physics, Boltzmanngasse 5, A-1090 Vienna, Austria}

\author{Daris Samart}
\email{darisa@kku.ac.th}
\affiliation{Khon Kaen Particle Physics and Cosmology Theory Group (KKPaCT), Department of Physics, Faculty of Science, Khon Kaen University,123 Mitraphap Rd., Khon Kaen, 40002, Thailand}

\begin{abstract}
In strongly interacting massive particle (SIMP) scenarios, dark matter is comprised of stable dark pions whose $3\to 2$ or $4\to 2$ reactions set the dark matter relic abundance. Recent work has shown that shallow two-pion bound states significantly affect the freeze-out, but did not establish whether such states actually form. In this work we demonstrate that a scalar isosinglet bound state does exist in a well-defined region of parameter space by solving an on-shell Lippmann--Schwinger equation in a chiral-unitary framework and analyzing the $S$-wave $\pi\pi$ amplitude in the complex energy plane. We determine the range of $m_\pi/f_\pi$ for which a pole appears below the two-pion threshold, extract the corresponding residue, and, in the non-relativistic limit, obtain the bound-state wave function at the origin, $|\Psi(0)|$, which controls bound-state-assisted annihilation and decay rates relevant for catalyzed freeze-out. Comparing this T-matrix based result with variational estimates using simple finite-range potentials, we find agreement within order-one factors for shallow binding. For binding energies of order the freeze-out temperature, $E_B \sim m_\pi/20$, we obtain $|\Psi(0)|\sim \mathcal{O}(0.1)\,m_\pi^{3/2}$, thereby supporting the parametric assumptions used in previous phenomenological analyses.
\end{abstract}

\maketitle

Strongly self-interacting dark matter (DM) with scattering cross section over DM mass at the level of 
$\sigma_{\rm self}/m_{\rm DM}\sim 0.1$–$10~{\rm cm^2/g}$
has become an appealing scenario,  modifying small-scale astrophysical structures relative to the collisionless DM  framework~\cite{Gradwohl:1992ue, Spergel:1999mh}. This can ease tensions associated with central densities and inner slopes in dwarf galaxies and galaxy clusters, while remaining compatible with large-scale structure and cluster constraints. 
In this context, the so-called strongly interacting massive particle (SIMP) mechanism~\cite{Hochberg:2014dra} provides a concrete thermal production scenario in which number-changing $3\to 2$ or $4\to 2$ processes among dark-sector states set the observed relic abundance, rather than the usual $2\to 2$ DM annihilation into Standard Model (SM) particles. In the SIMP scenario, the same interactions naturally give rise to sizable elastic self-scattering, thereby tying together the DM relic density and small-scale structure phenomenology. This framework has received substantial attention in the past, see, e.g.~\cite{Hochberg:2015vrg, Kuflik:2015isi,Bernal:2015bla, Bernal:2015xba, Bernal:2015ova,  Choi:2015bya, Choi:2016hid,Soni:2016gzf, Kamada:2016ois,  Bernal:2017mqb,Cline:2017tka, Choi:2017mkk,  Kuflik:2017iqs, Heikinheimo:2018esa, Choi:2018iit,Hochberg:2018rjs, Bernal:2019uqr, Choi:2019zeb,  Katz:2020ywn, Smirnov:2020zwf, Xing:2021pkb, Braat:2023fhn, Bernreuther:2023kcg,Garcia-Cely:2024ivo,Alfano:2025non,Garcia-Cely:2025flv}.

A particularly economical class of SIMP models identifies the DM particle with pseudo-Nambu-Goldstone bosons $\pi$ (``dark pions'') of a new confining gauge sector. Prominent ultraviolet\,(UV)-complete confining theories feature gauge groups $\mathrm{Sp}(2N_c)$ or $\mathrm{SU}(N_c)$ with a small number of light fermion flavors.
In these realizations, the leading $3\to 2$ processes are generated by the Wess–Zumino–Witten (WZW) term~\cite{Wess:1971yu,Witten:1983tw}, while $2\to 2$ self-scattering arises from the usual chiral Lagrangian interactions. In those explicit constructions based on confining gauge theories, reproducing the observed relic abundance typically requires~\cite{Hochberg:2014kqa}
\begin{equation}
  {m_\pi}/{f_\pi} \sim \mathcal{O}(\text{few})\,,
\end{equation}
where $m_\pi$ is the dark pion mass and $f_\pi$ is the pion decay constant, so that the chiral expansion is pushed towards the boundary of its nominal regime of validity.
This tension has motivated systematic analyses of strongly coupled SIMP theories~\cite{Kulkarni:2022bvh}, and consideration of higher-order terms in the chiral Lagrangians~\cite{Hansen:2015yaa}.
Lattice simulations of $\mathrm{Sp}(4)$ and related theories~\cite{Pica:2016zst,Bennett:2017kga,Bennett:2019jzz,Bennett:2019cxd,Bennett:2022yfa,Bennett:2023wjw} have begun to map out the light spectrum and extract low-energy constants, providing first-principle information on the relation between $m_\pi$ and $f_\pi$, and phenomenologically relevant quantities~\cite{Kulkarni:2022bvh,Dengler:2024maq,Dengler:2025ulb}.

Dark confining theories generically contain a tower of composite states beyond the lightest pseudoscalars, including scalar and vector mesons, with some works addressing their potential impact~\cite{Choi:2018iit,Berlin:2018tvf,Bernreuther:2023kcg}. Recently, Ref.~\cite{Chu:2024rrv} emphasized that shallow two-pion bound states can significantly affect SIMP freeze-out in dark-pion models. In particular, an $S$-wave scalar bound state $X=[\pi\pi]$ with its binding energy of the order of the freeze-out temperature, $E_B\sim T_{\rm fo}$, can mediate efficient bound-state-assisted annihilation processes, thereby modifying both the relic abundance and the late-time self-interaction phenomenology. However, that analysis treated the existence of such a state and its wave function at the origin, $|\Psi(0)|$, as input, constrained only by general considerations on short-range potentials. From the point of view of a UV-complete SIMP model based on an $\mathrm{Sp}(2N_c)$ gauge theory, it remains a dynamical question whether such a state actually forms for cosmologically relevant values of $m_\pi/f_\pi$, and, if so, what range of $|\Psi(0)|$ is realized.
The prospect that a scalar state appears close to the two-pion threshold and can behave as a shallow two-pion molecule once the interactions become sufficiently strong has a potential SM analogue: as reviewed in~\cite{Albaladejo:2012te, Guo:2017jvc}, for $m_\pi \gtrsim 300$–$500$\,MeV, it is generally believed that the SM sigma meson ($f_0(500)$) can be viewed as a hadronic molecule just below the two-pion threshold. The existence of an analogous dark scalar in $\mathrm{Sp}(4)$ theories has also been suggested in the lattice literature~\cite{Drach:2021uhl,Kondo:2022lgg}, but a quantitative study of the conditions for its formation and of its internal structure in the SIMP-relevant parameter range has been lacking.

In this work we address this question quantitatively in a concrete SIMP-motivated setup, using methods developed for SM strong dynamics. We consider a QCD-like theory with $N_c$ dark colors in the fundamental representation of $\mathrm{Sp}(4)$ and $N_f=2$ Dirac fermions, so that the global symmetry breaking pattern is $\mathrm{SU}(4)\to\mathrm{Sp}(4)$ and the dark pions form a two-index anti-symmetric representation, denoted by ${\bf 5}$, of $\mathrm{Sp}(4)$. Working within the chiral effective field theory of these dark pions, and using the $\mathcal{O}(p^2)$ and $\mathcal{O}(p^4)$ results of Refs.~\cite{Bijnens:2011fm,Hansen:2015yaa}, we study two-pion scattering in a chiral-unitary framework. We determine the range of $m_\pi/f_\pi$ for which a pole in the singlet $S$-wave T-matrix moves below the two-pion threshold when the amplitude is unitarized in a Lippmann-Schwinger approach, and extract the residue at the pole to obtain the non-relativistic bound-state wave function at the origin, $|\Psi(0)|$, in a controlled way. This turns $|\Psi(0)|$ from a free phenomenological parameter into a prediction of the underlying confining dynamics, directly relevant for bound-state–assisted annihilation in catalyzed SIMP freeze-out. We also compute the $P$-wave scattering amplitude and assess whether an $\ell=1$ two-pion bound state is viable within the regime where the chiral-unitary treatment remains reliable.

The paper is organized as follows. In Sec.~\ref{ch1} we review the SIMP dark-pion framework based on an $\mathrm{Sp}(4)$ confining gauge theory with global symmetry breaking $\mathrm{SU}(4)\to\mathrm{Sp}(4)$, summarize the relevant features of the chiral effective theory, and specify the parameter choices motivated by lattice results. In Sec.~\ref{ch2} we construct the two-pion scattering amplitudes in the singlet $S$-wave and in the relevant $P$-wave channels, project them onto partial waves, and implement unitarization in a Lippmann-Schwinger framework. Sections~\ref{ch3} and~\ref{ch4} contain our main results: in Sec.~\ref{ch3} we determine the range of $m_\pi/f_\pi$ for which a scalar singlet pole moves below threshold, quantify its binding energy; we also analyze the $P$-wave amplitude and assess the viability of an $\ell=1$ bound state. In Sec.~\ref{ch4} we extract the corresponding non-relativistic wave function of the scalar bound state at the origin $|\Psi(0)|$, and discuss the implications of our findings for bound-state–assisted SIMP freeze-out and late-time self-interactions, emphasizing how $|\Psi(0)|$ and the existence of the scalar bound state feed into phenomenological analyses. We conclude in Sec.~\ref{ch5} and collect technical details of the amplitude projections, unitarization procedure, and alternative methods in several appendices.

\section{The dark sector framework} 
\label{ch1}

Throughout this work, we consider a QCD-like theory with $N_c$ dark colors, transforming in the fundamental representation of the $\mathrm{Sp}(N_c)$ gauge group. The matter content contains $N_f$ Dirac fermions, or, equivalently, $2N_f$ Weyl fermions  (``quarks''). The pseudo-real nature of color $\mathrm{Sp}(N_c)$ implies a global  symmetry $G = \mathrm{SU}(2N_f)$ acting on the massless Weyl fermions. Nevertheless, confinement (or a quark mass term) reduces the global symmetry to $H = \mathrm{Sp}(2N_f)$ and thus generates $(2N_f^2 - N_f - 1)$ dark pions, or pseudo-Goldstone bosons, after color confinement. For simplicity, these fermions are taken to have equal and non-vanishing masses $m_q$, leading to a common pion mass $m_\pi$. As is well known, this setup allows for an odd-numbered Wess-Zumino-Witten (WZW) interaction among these dark pions~\cite{Hochberg:2014kqa}.

By labeling the broken group generators by $X^a$ and unbroken ones by $T^a$ ($a=1,\dots, 2N_f^2 - N_f - 1$), we obtain the associated Goldstone boson manifold $G / H$, with a coset representative
\begin{eqnarray}
u=\exp \left(\frac{i}{\sqrt{2} f_\pi} X^a \pi^a\right)\,,
\end{eqnarray}
where the generators are normalized to $\left\langle X^a X^b\right\rangle=\delta^{a b}$ and $\langle \,.\,\rangle$ denotes trace in flavor space and a summation over repeated indices is implied. The quantity $u$ transforms under $\mathrm{SU}(2N_f)$ as
\begin{eqnarray}
u ~ \rightarrow ~ g u h^{\dagger} \,,
\end{eqnarray}
with $g \in G$ and the compensator field $h \in H$. %
In this setup, the orientation  and the associated Goldstone fluctuations of the chiral condensate are characterized by 
\begin{equation}
\Sigma_c = J_A   \equiv   \begin{pmatrix}
0 & \text{I} \\
-\text{I} & 0 
\end{pmatrix}   \,,\qquad     \Sigma = u \,\Sigma_c  u^\text{T} = u \,J_A  u^\text{T} \,,
\end{equation}
respectively, where $\Sigma$ transforms under $\mathrm{SU}(2N_f)$ as
\begin{eqnarray}
\Sigma ~ \rightarrow ~ g \, \Sigma \,  g^\text{T} \,.
\end{eqnarray} 
The relations among these $\mathrm{SU}(2N_f)$ generators,  $T^a J_A = - J_A (T^a)^\text{T}$ and $X^a J_A = J_A (X^a)^\text{T}$, ensure that  $ \Sigma $ is invariant if $g \in H$, while it is moved along the $G/H$ manifold for transformations induced by the broken generators.
Throughout this work, we use $u$ to write down the chiral effective Lagrangian, which  is sometimes referred to as the Callan-Coleman-Wess-Zumino scheme or construction~\cite{Coleman:1969sm,Callan:1969sn}.  For the description of the Lagrangian using $\Sigma$ instead, see e.g. \cite{Gasser:1984gg, Kogut:2000ek}.

Concretely, we follow \cite{Bijnens:2011fm, Hansen:2015yaa} and organize the effective  Lagrangian in the standard chiral expansion $\mathcal{L}_{\rm eff} = \mathcal{L}_2 + \mathcal{L}_4 + \dots$,  where $\mathcal{L}_{2 k}$ collects all operators of chiral order $\mathcal O\left(p^{2 k}\right)$.
The $\mathcal{O}(p^2)$ and $\mathcal{O}(p^4)$ interactions are concretely given by~\cite{Gasser:1983yg},
\begin{align}
\label{L2}
\mathcal{L}_2& =\frac{f_\pi^2}{4}\left\langle u_\mu u^\mu+\chi_{+}\right\rangle,
\\
\label{L4}
\mathcal{L}_4& = L_0\left\langle u_\mu u_\nu u^\mu u^\nu\right\rangle+L_1\left\langle u_\mu u^\mu\right\rangle\left\langle u_\nu u^\nu\right\rangle  
+L_2\left\langle u_\mu u_\nu\right\rangle\left\langle u^\mu u^\nu\right\rangle+L_3\left\langle u_\mu u^\mu u_\nu u^\nu\right\rangle \notag \\
& +L_4\left\langle u^\mu u_\mu\right\rangle\left\langle\chi_{+}\right\rangle+L_5\left\langle u^\mu u_\mu \chi_{+}\right\rangle  +L_6\left\langle\chi_{+}\right\rangle^2+L_7\left\langle\chi_{-}\right\rangle^2+\frac{1}{2} L_8\left\langle\chi_{+}^2+\chi_{-}^2\right\rangle ,
\end{align}
where the bi-fields are defined as  
\begin{equation}
u_\mu=i\left( u^{\dagger} \partial_\mu u -u \partial_\mu    u^{\dagger} \right) \,, ~~~~
\chi_{ \pm}=u^{\dagger} (\chi J_A) u^{\dagger} \pm u (J_A \chi^\dagger)  u\,,\qquad  \chi = 2B_0 
\begin{pmatrix}
0 & \mathcal M_q \\
-\mathcal M_q   & 0 
\end{pmatrix}  \,. 
\end{equation}
Here $f_\pi$ and $B_0$ are the leading order low-energy constants (LECs) with $f_\pi$ the chiral-limit dark pion decay constant and $B_0$ being proportional to the quark condensate; $\mathcal M_q$ is the quark mass matrix, and, under our equal mass assumption, $\mathcal{M}_q = m_q\,\mathbf{1}$ and hence $\chi = 2B_0 m_q J_A$. 
Above, $p$ is a generic soft-scale (external momenta and pseudo–Goldstone masses $m_\pi$) and $\mathcal O(p^{2k})$ is shorthand for powers of the dimensionless ratio $p/\Lambda$ with $\Lambda = 4\pi f_\pi$; in the small quark mass limit, one counts $\partial_\mu \sim O(p), m_q \sim O\left(p^2\right)$, and $m_\pi \sim O\left(p\right)$. 
For the LECs in the $\mathcal O(p^4)$ Lagrangian, we  use the modified $\overline{\mathrm{MS}}$ scheme, where  
\begin{equation}
L_i=L_i^r-\frac{\Gamma_i}{32 \pi^2} R\,,\quad {\rm with} \quad R=\frac{2}{\epsilon}+\log (4 \pi)-\gamma_E+1\,. 
\end{equation}
Here $\epsilon=4-d$, $\gamma_E=-\Gamma^{\prime}(1)$ is the Euler-Mascheroni constant; the constants $\Gamma_i$ can be found in Tab.~1 of Ref.\,\cite{Bijnens:2009qm}. The renormalized coefficients $L_i^r$ depend on the energy scale $\mu$ introduced by dimensional regularization. 
In this work, we adopt the central values of LECs from Ref.\,\cite{Bijnens:2014lea}, but rescale them down by one order of magnitude, to values of order $\mathcal{O}\big( 10^{-4}\big)$, in line with the expectations in \cite{Hansen:2015yaa}. Restricting ourselves to~\eqref{L2} and~\eqref{L4}, we have neglected  gauge or Yukawa interactions with external particles.  In this case, it is straightforward to see that the constructed Lagrangian is invariant under the global symmetry~$H$. 

Below, we simplify our analysis by adopting $N_f = 2$, which is a commonly used benchmark value in SIMP models. In fact, for $\pi\pi$ scattering in the flavor-singlet channel---which we will show is the most relevant one in the context of bound states---the corresponding scattering amplitude grows for increasing~$N_f$~\cite{Bijnens:2011fm}. For instance, at LO the scattering length increases linearly with $N_f$. Therefore, $S$-wave bound states are expected to form  with even smaller values of $m_\pi/f_\pi$, when adopting larger values of $N_f$.

\section{\boldmath$S$-Wave Scattering of Dark Pions}
\label{ch2}

This section lays the groundwork for the analysis of the $S$-wave pole by first providing the explicit expressions for the $\pi\pi$ scattering amplitudes. 
Such a pole  in the scattering amplitude with the center-of-mass energy $\sqrt{s} < 2m_\pi$, if present and physical, would indicate a two-pion bound state. 
As the pole approaches the two-pion threshold from below, the binding energy of the corresponding bound state decreases and eventually vanishes.
For $S$-wave elastic scattering, the low-momentum behavior can be parametrized in terms of the scattering length $a_S$ by the scattering phase shift
$
\delta_S(k) \simeq-k a_S$ for $k \rightarrow 0$ and $|k a_S| \ll 1
$.
In the usual convention, a weakly attractive interaction that is not strong enough to form a bound state has a negative scattering length, $a_S<0$, and hence a small positive phase shift, $\delta_S(k)>0$, or, equivalently, a positive scattering amplitude $\mathcal{M}>0$. As the attractive coupling is increased, $\left|a_S\right|$ grows. If 
one reaches 
\begin{equation}
  \delta_S \to {\pi \over 2}~\text{~~and~~}~a_S \to -\infty\,,  
\end{equation}
a zero-energy bound state appears as a genuinely non-perturbative effect.
An infinite scattering length corresponds to a pole in the scattering amplitude $\mathcal{M}$, or more precisely, in the partial wave projection of the T-matrix amplitude. Here we define the T-matrix operator from the S-matrix operator via $$\hat{\mathcal{S}} =  1 + i \hat{\mathcal{T}}\,,$$ where a  positive sign is chosen to match the most widely used convention, while Refs.~\cite{Oller:2000ma, Hyodo:2011ur} adopt a negative sign.\footnote{Others, e.g.,~\cite{Bijnens:2011fm,Oller:2024lrk}, while choosing a positive sign for $\hat{\mathcal{T}}$, define the scattering length with an opposite sign with respect to ours.}  
Thus, by locating zeros in the real part of the inverse T-matrix amplitude in a given partial wave, at a centre-of-mass energy $s < 4m^2_\pi$, 
one can diagnose whether $S$-wave or higher-$\ell$ bound states are expected to form. %

\begin{figure}[tb]
\centering
\includegraphics[width=0.8\columnwidth]{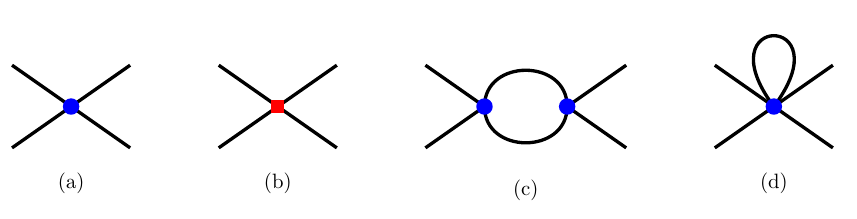}
\caption{The LO (a) and NLO (b--d) diagrams for pion-pion scattering. The solid blue circle is a vertex from $\mathcal{L}_2$, and the solid red square is a vertex from $\mathcal{L}_4$\,.
\label{fig:diagram2}}
\end{figure}

In what follows we analyze the T-matrix  organized in the chiral expansion in the vicinity of the two-pion threshold. The expansion parameter $p/\Lambda$ is hence $m_\pi/(4\pi f_\pi)$ as all external momenta are soft. The leading-order (\LO) $\mathcal{O}(p^2)$ contribution to the T-matrix element corresponds to the leftmost diagram~(a) shown in Fig.~\ref{fig:diagram2} and is solely due to~$\mathcal{L}_2$. The next-to-leading order (NLO) $\mathcal{O}(p^4)$ contribution arise due to the tree-level diagram~(b) from $\mathcal{L}_4$ and one-loop diagrams (c) and (d) with interactions from $\mathcal{L}_2$. We will estimate the associated theoretical uncertainty by comparing the contributions at each order. We emphasize that issues of unitarization at high energies---relevant when amplitudes from higher-dimensional operators grow with~$s$---are not a serious concern here, since we restrict ourselves to the low-energy regime with $s \leq  4m^2_\pi$.

\subsection{LO amplitude in the flavor-singlet channel}

Expanding the effective Lagrangian $\mathcal{L}_2$ of Eq.~\eqref{L2} in the pion fields up to sixth order, one finds
\begin{align}\label{eq:Lang}
  \mathcal{L}_{\rm 2}  &
\simeq  \frac{1}{2} m_\pi^2 \sum_{a} \,\pi^a \pi^a - \frac{1}{12 f_{\pi}^{2}} \sum_{a,b}\left[ \pi^b \pi^b \,(\partial_\mu \pi^a)(\partial^\mu \pi^a) 
    - \pi^b (\partial_\mu \pi^b) \,\pi^a (\partial^\mu \pi^a) \right]
    + \frac{m_{\pi}^{2}}{48 f_{\pi}^{2}} \left(\sum_{b} \pi^b \pi^b \right)^2 \,
     \notag  \\
     & \quad 
     +\frac{1}{180 f_{\pi}^{4}} \sum_{a,b,c} \pi^a \pi^a\left[\pi^b\pi^b (\partial_\mu \pi^c)( \partial^\mu \pi^c) -\pi^b (\partial_\mu \pi^b)  \pi^c (\partial^\mu \pi^c) \right] - \frac{m_{\pi}^{2}}{2880 f_{\pi}^{4}} \left(\sum_{a}\pi^a \pi^a\right)^3\,.
\end{align}
Here and in the following, the sums over flavor indices run as $a,b,c= 1,\dots, 5$. 
At LO, considering the elastic $\pi^a$-$\pi^a$ scattering of same-flavor pions,  the last term of the first line produces a contributing four-point vertex and its positive coefficient gives an attractive force.\footnote{As is well known, the associated negative potential is not bounded from below, and additional potential terms are always needed.}
For its $S$-wave component, this corresponds to a positive scattering amplitude $\mathcal{M}_{aa\to aa}>0$, and a negative scattering length in the weak coupling limit.

To proceed systematically, we label the flavor indices and the momenta for incoming and outgoing particles in the general two-body process as
\begin{eqnarray}
\pi^{a_1}(p_1) + \pi^{a_2}(p_2) \to \pi^{a_3}(p_3) + \pi^{a_4}(p_4)\,.
\end{eqnarray}
As is usually done, we parameterize the kinematics in terms of the Mandelstam invariants $s =\left(p_1+p_2\right)^2,  t=\left(p_1-p_3\right)^2,$ and $ u=\left(p_1-p_4\right)^2$, which satisfy $s+t+u=\sum_{i=1}^4 p_i^2$. For elastic $\pi \pi$ scattering with all external pions on shell, $p_i^2=m_\pi^2$, this implies, as usual, $s+t+u=4 m_\pi^2$. In the physical scattering region these variables are real, but below we will regard $s, t, u$ as complex variables when discussing the analytic structure of the amplitudes.
At leading order one finds the following independent tree-level amplitudes:
\begin{eqnarray}
   \mathcal{M}^\text{LO}_{aa\to aa} & = & {3m_\pi^2 \over 6f_\pi^2} \,,\\
   \mathcal{M}^\text{LO}_{ab\to ab} & = & {- 3m_\pi^2  + 3t \over 6f_\pi^2}  \qquad (a\neq b)\,,\\
   \mathcal{M}^\text{LO}_{aa\to bb} & = &{- 3 m_\pi^2 + 3s \over 6f_\pi^2}  \qquad (a\neq b)\,,
\end{eqnarray}
where $a$ and $b$ label the pion flavors. The expressions 
suggest an attractive (a repulsive) force between identical (different) pions for non-relativistic scattering.  These directly correspond to the $S$-wave amplitudes by setting $s =4m_\pi^2$ and $t=u=0$. So, they can be used to calculate the partial-wave T-matrix at LO later. 

In the language of group theory, in the case of $N_f =2$, the four Weyl fermions (``quarks'') transform in the fundamental representation of~Sp(4), and the pions, as two-quark states, are in the adjoint antisymmetric representation, which can be written as ${\bf 5}$. As a result, the intermediate state of two-pion scattering $\pi^a \pi^b$ spans a tensor product $\mathbf{5} \otimes \mathbf{5}$, which can be de-composed into three irreducible Sp(4) multiplets as, 
\begin{eqnarray}
\label{decomposition}
{\bf 5}\; \otimes \; {\bf 5} = {\bf 1} \;\oplus {\bf 10} \;\oplus {\bf 14}\,.
\end{eqnarray}
The $\mathbf{1}$ is the flavor singlet (isoscalar) channel, while $\mathbf{10}$ and $\mathbf{14}$ are non-singlet ``isospin'' channels; for a detailed group theory discussion of Sp(4),  see App.~\ref{app:grouptheory}.  A convenient normalized basis state for the singlet in the two-pion sector is
\begin{equation}
   |  R_I \rangle  = \sum_{a = 1}^5\frac{1 }{\sqrt{5} } |\pi^a \rangle | \pi^a \rangle  \,, 
\end{equation} 
which, as the name suggests, is invariant under Sp(4) flavor rotations.

It is useful to regard the scattering amplitude as an T-matrix  operator $\hat{\mathcal{T}}$ acting on the two-pion flavor space, with matrix elements $\langle\pi^{a_3} \pi^{a_4}| \hat{\mathcal{T}}|\pi^{a_1} \pi^{a_2}\rangle $. 
Projecting this operator onto the singlet state defines the isosinglet amplitude, $ \mathcal{M}_I(s, t, u)$, via
\begin{equation}\label{eq:TonRI}
\left \langle R_I \right| \hat{\mathcal{T}}\left|R_I \right\rangle  =  (2\pi)^4\delta^{(4)}(p_1 + p_2 - p_3 -p_4) \mathcal{M}_I(s, t, u)  \, . 
\end{equation}
At leading order, $\mathcal{M}_I(s, t, u)$ is thus given by a specific linear combination of the channel amplitudes $\mathcal{M}_{a a \rightarrow a a}^{\mathrm{LO}}, \mathcal{M}_{a b \rightarrow a b}^{\mathrm{LO}}$, and $\mathcal{M}_{a a \rightarrow b b}^{\mathrm{LO}}$ above. Consequently, one obtains its leading-order expression as 
\begin{equation}\label{eq:swaveLOM}
\mathcal{M}^{\mathrm{LO}}_I(s, t, u)    =  {4s - 3m_\pi^2 \over 2f_\pi^2}\,, 
\end{equation}
in agreement with Eq.~(69) of Ref.~\cite{Bijnens:2011fm}.  Near threshold, $s = 4m_\pi^2$, the amplitude is positive, suggesting a desired attractive potential. 
By contrast, projecting the T-matrix operator $\hat{\mathcal{T}}$ at leading order onto the $\mathbf{10}$ and $\mathbf{14}$ irreducible representations yields amplitudes dominated by the mixed-flavor channels with $a \neq b$. Near threshold these combinations are negative (or at best small), corresponding to repulsive or only weakly attractive interactions. Therefore, in what follows we focus on the singlet channel, which is the most attractive $S$-wave channel and thus the most promising for forming a two-pion bound state.

\subsection{LO, NLO (and NNLO) amplitudes in the flavor-singlet channel}

For the singlet channel,  we include both LO and NLO amplitudes, making extensive use of the  scattering amplitudes computed in~\cite{Bijnens:2011fm,Hansen:2015yaa}.  
The scattering amplitudes,  up to NLO, may be cast in the following form~\cite{Hansen:2015yaa}
\begin{equation}
\begin{aligned}
{\mathcal M} (s, t, u) 
=  & \xi^{a_1 a_2 a_3 a_4} \mathcal{B}(s, t, u)+\xi^{a_1 a_3 a_2 a_4} \mathcal{B}(t, u, s) 
+\xi^{a_1 a_4 a_2 a_3} \mathcal{B}(u, s, t)\\
&+ \delta^{a_1 a_2} \delta^{a_3 a_4} \mathcal{C}(s, t, u) 
+\delta^{a_1 a_3} \delta^{a_2 a_4} \mathcal{C}(t, u, s)+\delta^{a_1 a_4} \delta^{a_2 a_3} \mathcal{C}(u, s, t)\,,
\end{aligned}
\label{general-amp}
\end{equation}
where  the factor $\xi^{a_1a_2a_3a_4}$ stems from the $\mathrm{SU}(4) \rightarrow S p(4)$ pattern of chiral symmetry breaking,  given by
\begin{eqnarray}
\xi^{a_1 a_2 a_3 a_4}=\frac{1}{2}\left(\delta^{a_1 a_2} \delta^{a_3 a_4}-\delta^{a_1 a_3} \delta^{a_2 a_4}+\delta^{a_1 a_4} \delta^{a_2 a_3}\right),
\end{eqnarray}
with $\{a_1, a_2, a_3, a_4\}=1, \ldots, 5$. In addition,
 the $\mathcal{B}$ and $\mathcal{C}$ functions are defined by
\begin{eqnarray}
\mathcal{B}(s, t, u) &=& \mathcal{B}_{\LO}(s, t, u) + \mathcal{B}_{\NLO}(s, t, u)\,,
\\
\mathcal{C}(s, t, u) &=& \mathcal{C}_{\LO}(s, t, u) + \mathcal{C}_{\NLO}(s, t, u)\,,
\label{def-B-C}
\end{eqnarray}
with their full expressions given in App.~\ref{app:amplitudes}.\footnote{The physical pion mass and $f_\pi$ also receive subleading corrections from the NLO terms~\cite{Bijnens:2007yd, Kolesova:2025ghl}. These corrections are assumed to be absorbed into a re-definition of the LEC parameters, and are thereby accounted for within the associated uncertainties here.  }

Reference\,\cite{Bijnens:2011fm} classifies the two–meson intermediate states in the pseudo-real case $\mathrm{SU}\left(2 N_f\right) / \operatorname{Sp}\left(2N_f\right)$ in terms of $\operatorname{Sp}\left(2N_f\right)$ irreducible representations $R_X$ using Young tableaux. 
In the $\operatorname{Sp}(4)$ case ($N_f=2$) this reduces to the three relevant channels shown in~\eqref{decomposition}.
In the notation of~\cite{Bijnens:2011fm} the singlet $\mathbf{1}$, the 10-plet $\mathbf{10}$ and the 14-plet $\mathbf{14}$ representation are denoted by their definite flavor-symmetry properties as  $R_I, R_S$ and $R_{M S}$, respectively. Here, the label S refers to a symmetric and MS to a mixed-symmetric representation.\footnote{$R_S$ is in fact totally anti-symmetric under the exchange of the two initial states~\cite{Bijnens:2011fm}. }
To obtain the amplitude in each of these channels, one projects the full flavor amplitude onto the corresponding irreducible representations using the projection operators given in Tab.~4 of~\cite{Bijnens:2011fm}.
Concretely for Sp(4), the resulting invariant amplitudes read
\begin{eqnarray}
\mathcal{M}_{I}(s,t,u) &=& \frac52\,\Big[ \mathcal{B}(s,t,u) + \mathcal{B}(t,u,s) - \frac{3}{5}\,\mathcal{B}(u,s,t) \Big] + 5\,\mathcal{C}(s,t,u) + \mathcal{C}(t,u,s) + \mathcal{C}(u,s,t)\,,
\label{amp-1}
\\
\mathcal{M}_{S}(s,t,u) &=& \mathcal{B}(t,u,s) - \mathcal{B}(s,t,u) + \mathcal{C}(t,u,s) - \mathcal{C}(u,s,t)\,,
\label{amp-10}
\\
\mathcal{M}_{MS}(s,t,u) &=& \mathcal{B}(u,s,t) + \mathcal{C}(t,u,s) + \mathcal{C}(u,s,t)\,.
\end{eqnarray}
As already discussed at the end of the previous subsection, $\mathcal{M}_{I}$ is the channel of interest in this work, since it is the most attractive $S$-wave channel.
At LO and NLO levels, the $S$-wave projection of its corresponding amplitudes are given by:
\begin{eqnarray}
 T_{\LO}(s)~ \equiv ~ T^{S}_{\LO,\,I}(s) &=& \frac12\frac12\int_{-1}^{+1} d(\cos\theta)\,P_0(\cos\theta)\,\mathcal{M}^{\LO~}_{I} 
\,, \label{vLO-14}
\\
 T_{\NLO}(s)  ~\equiv   ~ T^{S}_{\NLO,\,I}(s)  &=& \frac12\frac12\int_{-1}^{+1} d(\cos\theta)\,P_0(\cos\theta)\,\mathcal{M}^{\NLO}_{I} 
\,,\label{eq:TNLO}
\end{eqnarray}
where the integral is over the cosine of the center-of-mass scattering angle, $\cos\theta$; of course, $P_0(\cos\theta) =1$.
One factor of one-half accounts for the $S$-wave projection normalization and another factor of one-half is for identical particles in both,  initial and final states~\cite{Oller:1998zr}. As already implied in the equations, unless otherwise stated, for simplicity we omit the superscript ``${S}$'' for the $S$-wave projection and subscript ``$I$'' for the flavor singlet channel, as this is the main focus of our work. To be explicit,  in  the flavor singlet channel, we use $T(s)$ and $v(s)$ for the $S$-wave projection of scattering amplitude and  interaction kernel, respectively. A subscript, such as ``$\LO$'' and ``$\NLO$'', is added to indicate contributions from leading order or next-to-leading order. 

\begin{figure}[tb]
\includegraphics[width=8.5cm,height=6.1cm]%
{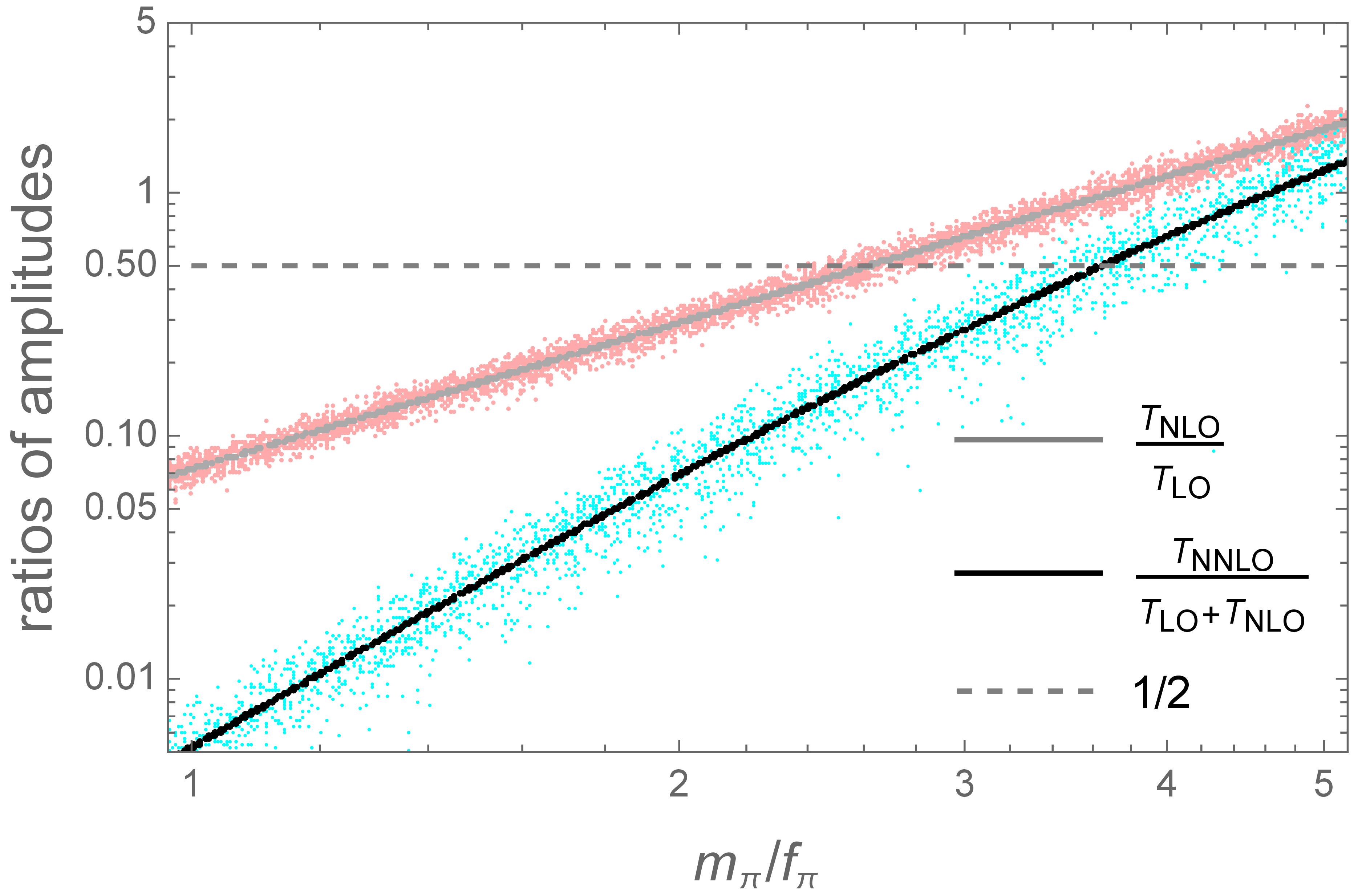}~~
\includegraphics[width=8.5cm,height=6cm]{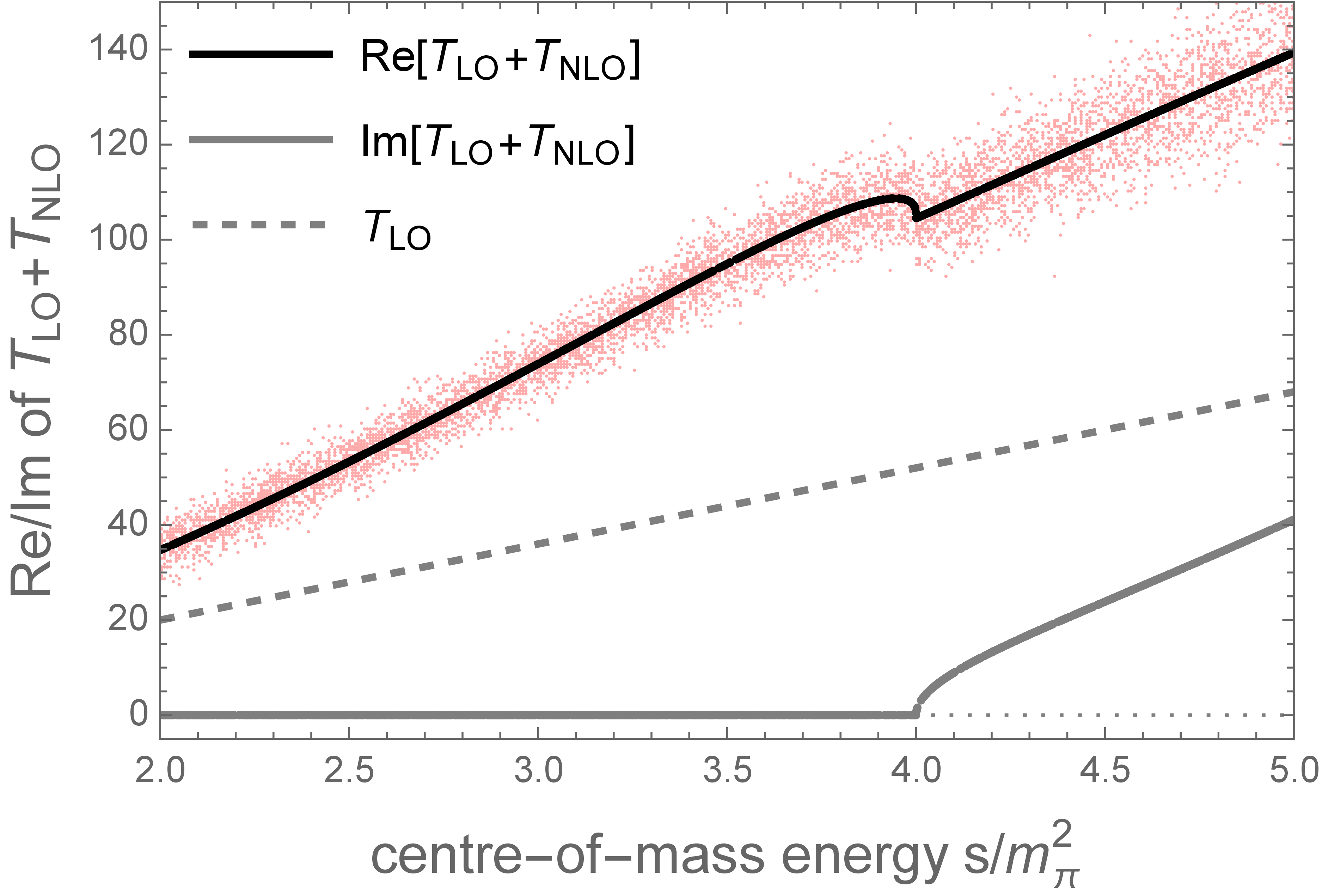}
\caption{{\it Left:} Comparative contributions of  NLO and NNLO T-matrix amplitudes relative to the one of lower chiral order, after taking $S$-wave projection  for  the flavor singlet channel at $s= (1.95m_\pi)^2$. {\it Right:} Real and imaginary components  of the amplitudes at LO and NLO level, as a function of $s$ for $m_\pi/f_\pi =4$. In all figures, uncertainties of LECs at NLO (combination of NLO and NNLO) are included by adding independent Gaussian fluctuations to their central values with a standard deviation of $10^{-4}$  ($10^{-6}$),  shown as pink (cyan) dots. }
\label{fig:perturbativity}
\end{figure}

 We now compare the contributions to the scattering amplitude of  the flavor singlet channel at different orders after the $S$-wave projection, in order to estimate the range of effective couplings,  $m_\pi/f_\pi$,  for which the perturbative expansion is valid, determining where the LO and NLO results can be trusted (for $s <  4m_\pi^2$) .
In the numerical analysis, we account for the uncertainties of the LECs by assigning to each of them an independent Gaussian fluctuation with a common standard deviation of $10^{-4}$  at NLO and $10^{-6}$ at NNLO level.
The resulting ratios are shown in the left panel of Fig.~\ref{fig:perturbativity}, indicating that the LO result is reliable only for $m_\pi/f_\pi \lesssim  2.6$, while the sum of LO and NLO terms dominates the total amplitude up to $m_\pi/f_\pi \gtrsim  4$. %
 For the numerical results,  we fix the dimensional regularization scale $\mu = 2m_\pi$ and the cutoff momentum $q_\text{max} =3m_\pi$, and evaluate the ratios at $s =(1.95m_\pi)^2$.   The bound states obtained below typically locates at smaller $s$, where higher-order contributions are even less important. 
 Right panel of Fig.~\ref{fig:perturbativity}  in turn illustrates for a fixed value of $m_\pi/f_\pi$ the dependence of the corresponding $S$-wave projected amplitudes, $T_\text{LO}$ and $T_\text{NLO}$, on $s$ in the vicinity of the $4m_\pi^2$ threshold. At tree level, the amplitude is always real, and increases with the centre-of-mass energy. At loop level, the imaginary part of the amplitude appears for $s$ above the threshold, as required by the optical theorem. This imaginary component does not rely on the LECs, as can  be seen from the optical theorem as well. Moreover, it is relatively small compared with the real part of the total amplitude, $T_{\LO} + T_{\NLO}$, having at most a mild effect on the numerical results, even at $s> 4m_\pi^2$.

Consequently, we always adopt the LO and NLO combination below, and  consider  $m_\pi/f_\pi \lesssim 5$, where higher-order contributions are subleading. For even larger couplings, NNLO contributions would become dominant, see e.g.~\cite{Kulkarni:2022bvh}.\footnote{Interestingly, Ref.~\cite{Arthur:2014zda} suggests that the LO contribution alone can provide a good fit for certain scattering channel, even for $m_\pi/f_\pi \sim 10$.} 
\section{Bound States from \boldmath$T$-Matrix Poles}
\label{ch3}

In this section, we will adopt a unitary, on-shell Lippmann-Schwinger equation by using the effective scattering potential that we  derive from the LO and NLO scattering amplitudes above. We will then search for the pole positions in the unitarized  amplitude after partial wave projection, representing the bound state in the dark pion-pion system.

\subsection{The \boldmath$S$-wave case}

In the chiral unitary approach~\cite{Oller:1997ti,Oset:1997it,Oller:1998hw} one solves an on-shell–factorized Lippmann–Schwinger equation, where the interaction kernel is taken from the chiral Lagrangian, where the external legs are on shell and its iteration is encoded in a two-pion loop function $G(s)$ defined via an on-shell subtraction in the dispersion–relation framework. This construction ensures exact elastic unitarity of the partial-wave T-matrix and leads to a factorization of the transition amplitude in the integral equation, so that the
Lippmann–Schwinger equation can be written in the compact form (for a review, see\,\cite{Oller:2000ma,Hyodo:2011ur}),
\begin{eqnarray}
T (s) = \frac{v(s)}{1 + v(s)\,G(s)}\,,
\label{LS-CHUA}
\end{eqnarray}
where the loop function of the two-pion  propagators, $G(s)$, has the form of 
\begin{equation}
G(s)
=  i \int \frac{d^4 q}{(2 \pi)^4} \frac{1}{q^2-m_{\pi}^2+i \epsilon} \frac{1}{(P-q)^2-m_{\pi}^2+i \epsilon}\,,
\end{equation}
and its details are further given in App.~\ref{app-loop-function}.  
As mentioned above, we have omitted the superscript ``${S}$'' and the subscript ``$I$'' in both $T(s)$ and $v(s)$ for the $S$-wave projected amplitudes of the flavor singlet channel studied here. 
Since $G(s)$ results in a negative real number below the two-pion threshold $s < 4m_\pi^2$, we expect a pole only for positive  effective scattering  kernel, being consistent with the discussion above from the viewpoint of the scattering length.

We follow the treatment in \cite{Albaladejo:2012te,Oller:2024lrk}, which assumes a perturbative hierarchy of the scattering kernel, $v(s) \simeq v_{\LO}(s) + v_{\NLO}(s) + ...$, to expand the above expression of the T-matrix amplitude in terms of 
\begin{eqnarray}
T (s) = v_{\LO}(s) + v_{\NLO}(s) - v^2_{\LO}(s) G(s) + ...\,.
\label{eq:CHUA}
\end{eqnarray}
Meanwhile, in the last section, we already obtained first- and second-order contributions to the $S$-wave projected amplitude as $T(s) \simeq T_{\LO}(s) +  T_{\NLO}(s)$.
A matching of the two expressions of $T(s)$ thus yields
\begin{eqnarray}
v_{\LO}(s) &=& T_{\LO}(s)\,,
\\
v_{\NLO}(s) &\simeq& T_{\NLO}(s) + T^2_{\LO}(s) G(s)\,. 
\end{eqnarray}

\begin{figure}[tb]
\includegraphics[width=8.7cm,height=6cm]{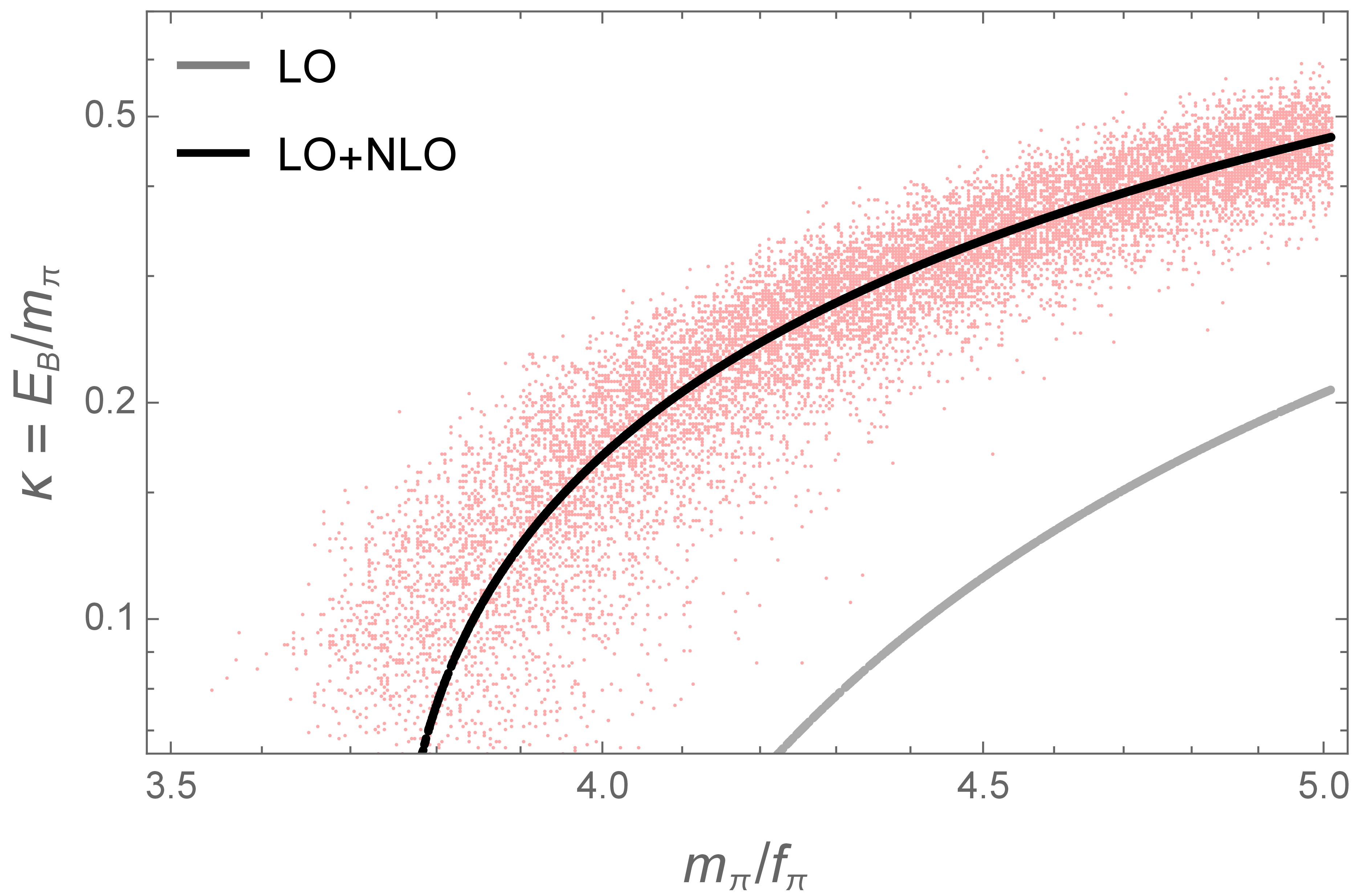}~~
\includegraphics[width=8.7cm,height=6cm]{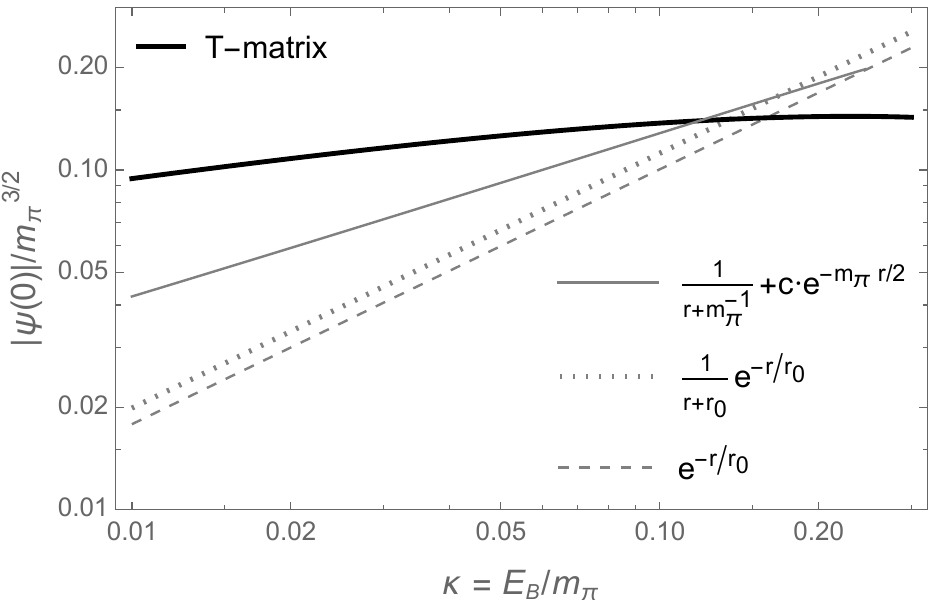}
\caption{{\it Left:} Binding energy $E_B$ normalized to pion mass $m_\pi$ of $S$-wave bound state as a function of effective coupling $m_\pi/f_\pi$. Pink dots show the range of variation under adding Gaussian noise  to the LECs, as specified in the main text. {\it Right:} Estimated absolute values of the bound state wave function at $r=0$ for small binding energies, calculated via the T-matrix and variational methods with various trial functions as labeled.
}
\label{fig:poleLoc}
\end{figure}

According to the right panel of Fig.~\ref{fig:perturbativity},  the real part of the $S$-wave projected amplitude is always  positive in the vicinity of $s\sim 4m_\pi^2$, suggesting an attractive force.     
And to search for the pole at the NLO level,  we need to calculate the solution of Mandelstam~$s$ below the $4m_\pi^2$ threshold\footnote{For a discussion of  other methods and reasons  why they do not perform well for our purpose here, see App.~\ref{App:methods}. } 
\begin{equation}
    1 + \left[ T_{\LO}(s) +  T_{\NLO}(s) + T^2_{\LO}(s) G(s)\right] G(s) =0\,.
\end{equation}
Denoting the pole solution as $s =s_P$, we show our  numerical results in the left panel of Fig.~\ref{fig:poleLoc}, where the $y$-axis shows the  pole location in terms of the (absolute value) of the binding energy $E_B$ normalized to the pion mass, $\kappa \,\equiv  \,E_B/m_\pi = (2m_\pi -\sqrt{s_P})/m_\pi$. Hence, a  positive $\kappa$ implies the formation of a physical bound state.  The plot suggest that a bound state appears when $m_\pi/f_\pi$ becomes slight larger than~3.5. As expected, the binding energy increases with the magnitude of the attractive coupling.  Bound states seem to appear already at LO level, for $m_\pi/f_\pi>4.2$, where, however, the convergence of the result cannot be guaranteed.

Such large values of the effective coupling, which are required to generate a $\pi\pi$ bound state in our analysis, correspond in the underlying confining theory to quark masses $m_q$ of order the confinement scale~$\Lambda$. In the chiral limit
with $m_q \ll \Lambda$ one has $m_\pi^2 f_\pi^2 \sim m_q \Lambda^3$ and $f_\pi \sim \mathcal{O}(\Lambda/4\pi)$, so that $m_\pi/f_\pi \propto \sqrt{m_q/\Lambda}$ is parametrically small and the chiral interactions are too weak to induce a bound state. Therefore, achieving $m_\pi/f_\pi \gtrsim \mathcal{O}(\text{few})$ requires $m_q \sim \Lambda$, i.e., pseudo-Nambu-Goldstone bosons whose masses are comparable to the confinement scale.
In this situation, the isosinglet state $|R_I\rangle$ resembles the sigma meson of QCD, if the physical pion mass were artificially increased above $\sim 400\,\text{MeV}$. Such a scalar can then be interpreted as a molecule composed of two pseudoscalar mesons; see, e.g.,
Refs.~\cite{Weinstein:1990gu,Baru:2003qq} for the kaon case, and
Refs.~\cite{Albaladejo:2012te,Pelaez:2015qba,Guo:2017jvc} for general reviews. Lattice results for the Sp(4) theory point to a similar picture when heavy quark (equivalently, heavy meson) masses are used~\cite{Drach:2021uhl,Bennett:2023rsl}.

For even larger values of $m_\pi/f_\pi$, a pole in the $S$-wave projected amplitude is still expected, but the corresponding state should behave more like a compact diquark--diantiquark configuration rather than a weakly bound two-pion molecule; see, e.g., Ref.~\cite{Eichmann:2015cra} and the reviews cited above. We emphasize that for such large values of $m_\pi/f_\pi$ theoretical uncertainties in both analytical and lattice calculations increase significantly, and quantitative results need to be interpreted with caution. 

\subsection{The \boldmath$P$-wave case} \label{sec:pwavePole}

In this subsection, we briefly study the $P$-wave channel to probe the strength of the corresponding interaction and assess whether an $\ell=1$ pole could exist. We note that only the amplitude for the $\pi^a\pi^b\to \pi^a\pi^b$\,($a\neq b$) channel contributes to the $P$-wave. At LO, its corresponding $P$-wave projected amplitude  reads
\begin{eqnarray}
   T^P_{{\LO},\,ab\to ab} (s) = \textcolor{black}{\frac12}\int_{-1}^{+1} d(\cos\theta) \, P_1(\cos\theta)\,\mathcal{M}_{\LO}  =   \frac{ m_\pi^2 }{3f_\pi^2}\left({s\over 4m_\pi^2} -1\right)\,, 
\end{eqnarray}
which is analogous to the SM pion case~\cite{Oller:2024lrk}. This results in negative values and thus repulsive forces below the $2m_\pi$ threshold, whereas above threshold, the interaction turns to be attractive. The absolute value of the loop function $G(s)$ is numerically small, with $|G(s)|\lesssim  0.023$   for  
$\sqrt{s} \leq 3m_\pi$. 
Achieving a pole in the unitarized amplitude with
\begin{equation}
    1 +   T^P_{\LO}(s)   G(s) =0\,,
\end{equation}
therefore requires $|T^{P}_{{\rm LO}}(s)| \gtrsim 43$ in the relevant $s$-range, corresponding to $m_\pi/f_\pi \gtrsim 10$. Such values are beyond the regime where the chiral expansion can be trusted: even after including NLO and NNLO contributions, higher-order corrections beyond NNLO are expected to be sizable. We thus conclude that, within the parameter region where our chiral-unitary framework is under quantitative control, the $P$-wave interaction is too weak to generate an $\ell=1$ pole.
This suggests that in contrast to the sigma meson, the rho meson cannot be explained as a bound state or resonance composed of two pions, even in the heavy quark case. 

Indeed, lattice results for the $\mathrm{Sp}(4)$ theory
\cite{Drach:2021uhl,Bennett:2023rsl}, as well as other studies of QCD-like theories~\cite{Maris:2005tt,Arthur:2016dir,Kulkarni:2022bvh,Alfano:2025non}, indicate that for $m_\pi/f_\pi$ larger than a few the lightest vector meson can become lighter than the sigma scalar and lie below the $2m_\pi$ threshold. In this regime the rho meson is a stable vector resonance and cannot be interpreted as the $\ell=1$ excited bound state of two pions within a purely pseudoscalar description. Quite the opposite: in the $P$-wave channel the {\it pre-existing} light rho-meson resonance must be included explicitly in the effective chiral Lagrangian, which substantially modifies the $P$-wave amplitude $T^P_{\LO}(s)$ compared to the simple contact interaction discussed above; see, e.g., Refs.~\cite{Bernard:1991zc,Oller:1998zr}. As a consequence, the location and nature of $P$-wave poles cannot be reliably inferred from perturbative $\pi\pi$ scattering amplitudes built only from pseudoscalar degrees of freedom. A consistent treatment requires either lattice input for the vector channel or an explicit resonance Lagrangian that incorporates the rho meson from the outset.

\section{Bound-State Wave Function}
\label{ch4}

In most SIMP realizations, the dark pions must scatter efficiently with lighter degrees of freedom, typically SM particles, in order to dissipate energy during their freeze-out to become cold dark matter. Once a two–pion bound state $X=[\pi\pi]$ exists, the same interactions typically destabilize $X$, and allow it to decay into those lighter states. In addition, bound states open up new number-changing channels such as $XX\to\pi\pi$ and $\pi X\to\pi\pi$, which can catalyze freeze–out by turning the original $3\to 2$ and $4\to 2$ reactions into effective two–body processes. Those aspects were the subject of~\cite{Chu:2024rrv}.

The decay and annihilation processes are governed by short-distance operators acting on the two constituents of~$X$. When the corresponding operators are local in the relative coordinate, the matrix elements factorize into a short–distance coefficient and the bound-state wave function at zero separation, so that the rates scale as $|\Psi(0)|^2$ (or as $|\nabla\Psi(0)|^2$ for $P$-wave states). In particular, the efficiency of the catalyzed freeze-out mechanism studied in~\cite{Chu:2024rrv} is controlled by ratios of such bound–state-assisted annihilation rates to scattering and breakup rates, and thus ultimately by the value of $|\Psi(0)|$.

In this section, we will  estimate the wave function of the bound state in the non-relativistic limit as a function of the binding energy, using two independent methods.  

\subsection{Bound-state wave function from the non-relativistic T-Matrix}

In the non-relativistic limit, the two-pion bound state can be described in the centre-of-mass frame by a relative wave function $\Psi(\vec r)$, which one may view as the wave function of one pion in the rest frame of the other. We define
the coordinate-space wave function in terms of the relative momentum eigenstates $|\vec p\,\rangle$ as
\begin{equation}
  \langle \vec r \mid \Psi \rangle \equiv \Psi(\vec r)
  = \int \frac{d^3 p}{(2\pi)^{3/2}}\, e^{i\vec p \cdot \vec r}\,
    \langle \vec p \mid \Psi \rangle\,,
\end{equation}
and our goal is to determine $\Psi(0)$ following the method of
Ref.~\cite{Gamermann:2009uq}.

The bound state satisfies the Schrödinger equation
$  (E - \hat H_0)\,|\Psi\rangle = \hat V\,|\Psi\rangle\,,$
with the full Hamiltonian operator $\hat H = \hat H_0 + \hat  V$,   where $\hat H_0$ is the free part (including inertial mass) and $\hat  V$ is the non-relativistic potential. In the momentum basis this gives
\begin{eqnarray}
\langle\vec{p} \mid \Psi\rangle=\int d^3 k \int d^3 k^{\prime}\left\langle\vec{p}\left|\frac{1}{E-\hat  H_0}\right| \vec{k}\right\rangle\left\langle\vec{k}|\hat  V| \vec{k}^{\prime}\right\rangle\left\langle\vec{k}^{\prime} \mid \Psi\right\rangle \,.
\label{eq:sch-rew}
\end{eqnarray}
For the $S$-wave component, we approximate the potential by a separable contact interaction with a sharp momentum cutoff for $k^{(\prime)} \equiv |\vec k^{(\prime)}|$,
\begin{equation}
  \langle\vec k|\hat V|\vec k^{\,\prime}\rangle
  = -\,\tilde v\,\Theta(\Lambda-k)\,\Theta(\Lambda-k^{\prime})\,,
\label{eq:potential-s}
\end{equation}
where $\Lambda$ is the cutoff and the minus sign is chosen such that $\tilde v$ and the T-matrix amplitude have the same sign at the pole.
The free resolvent in the two-pion centre-of-mass frame is
$
  \langle\vec p|(E-H_0)^{-1}| \vec k\rangle
  = \delta^{(3)}(\vec p-\vec k)\,
    (E-2m_\pi-{\vec p^{\,2}}/{m_\pi})^{-1},
$
where we used $m_1=m_2=m_\pi$ and reduced mass $\mu=m_\pi/2$, so that the
relative kinetic energy is $\vec p^{\,2}/(2\mu)=\vec p^{\,2}/m_\pi$. Inserting
this and Eq.~\eqref{eq:potential-s} into Eq.~\eqref{eq:sch-rew} yields
\begin{align}
  \langle\vec p \mid \Psi \rangle
  &= - \int d^3 k \int d^3 k^{\prime}\,
      \delta^{(3)}(\vec p-\vec k)\,
      \frac{1}{E-2m_\pi-\frac{\vec p^{\,2}}{m_\pi}}\,
      \tilde v\,\Theta(\Lambda-k)\,\Theta(\Lambda-k^{\prime})\,
      \langle\vec k^{\prime} \mid \Psi\rangle \nonumber\\[2mm]
  &= -\,\frac{\Theta(\Lambda-p)}{E-2m_\pi-\frac{\vec p^{\,2}}{m_\pi}}\,
      \tilde v \int d^3 k^{\prime}\,
      \Theta(\Lambda-k^{\prime})\,\langle\vec k^{\prime} \mid \Psi\rangle\,,
\label{eq:expandPsi}
\end{align}
where $p=|\vec p|$. Within this model, we see that $\langle\vec p \mid \Psi \rangle$ has only support for $p<\Lambda$, and vanishes for larger momenta.
Multiplying both sides by $\Theta(\Lambda-p)$, and integrating over $d^3p$, gives
\begin{equation}
  \int d^3p\ \Theta(\Lambda-p) \langle\vec p \mid \Psi \rangle = -\,\tilde v\,\tilde G(E)\, \int d^3 k^{\prime}\,
       \Theta(\Lambda-k^{\prime})\,\langle\vec k^{\prime} \mid \Psi\rangle,
  \quad \text{where}\quad
  \tilde G(E) \,\,\equiv  \, \int d^3 p\,
    \frac{\Theta(\Lambda-p)}{E-2m_\pi-\frac{\vec p^{\,2}}{m_\pi}}\,.
\end{equation}
A non-trivial bound-state solution  therefore requires
$  1 + \tilde v\,\tilde G(E) = 0$,  which is precisely the pole condition of the non-relativistic two-body Green's function. The function $\tilde G(E)$ is the non-relativistic limit of the relativistic loop function $G(s)$ up to an overall factor,
$  \tilde G(E) = 16\pi^3 m_\pi \sqrt{s}\, G(s)\,,$
and $\tilde v\,\tilde G(E) = v\,G(s)$, since both combinations must reproduce the same pole position.\footnote{The relativistic loop function $G(s)$ used here corresponds to $G^{\text{FT}}(s)$ in Ref.~\cite{Gamermann:2009uq}. See App.~\ref{app-loop-function} for further details.}

We now turn to the normalization condition of the bound-state wave function. In momentum space it reads
\begin{equation}\label{eq:normalPsi}
    1= \int d^3 p\,\big|\langle\vec{p} \mid \Psi\rangle\big|^2 =   \int d^3 p \, \left(\frac{\Theta(\Lambda-p)  }{E-2m_\pi-\frac{\vec{p}^2}{m_\pi}} \right)^2 \,   \tilde  v^2  \left| \int d^3 k^{\prime}  \Theta\left(\Lambda-k^{\prime}\right)  \left\langle\vec{k}^{\prime} \mid \Psi\right\rangle  \right|^2  \,,
\end{equation}
where in the second equality the first integral identifies with $-d \tilde G(E)/dE$. This identity is the key observation that  allows us to express the wave function in coordinate space  at the origin, since the latter can be expressed as
\begin{equation}
 \Psi(0)  \equiv   \langle \vec r =0 \mid \Psi \rangle 
  \simeq \frac{1}{(2\pi)^{3/2}}  \int d^3 p\,
  \Theta(\Lambda-p)\,\langle\vec p\mid \Psi \rangle \,.
\end{equation}
Concretely, replacing the first and second integrals on the right hand side of Eq.~\eqref{eq:normalPsi} with $-d \tilde G(E)/dE$ and $\Psi(0)$, and using the equality $  1 + \tilde v\,\tilde G(E) = 0$ at the pole,  one obtains 
\begin{equation}
  |\Psi(0)|   = \frac{|\tilde G(E)| }{(2\pi)^{3/2}}\,  
     \left| \frac{d\tilde G(E)}{dE}\right|^{-1/2}
  \quad \text{at}\quad E = 2m_\pi -E_B\,.
\end{equation}
Thus, the bound-state wave function at the origin can be expressed in terms of the loop function $\tilde G(E)$, evaluated at the bound–state energy $E = 2m_\pi -E_B$. We note in passing that the corresponding non-relativistic bound-state condition $  1 + \tilde v\,\tilde G(E) = 0$
is associated with the pole condition in the relativistic formulation $  1 +  v(s)\, G(s) = 0$ whose solution determines the pole position $s_P = (2m_\pi -E_B)^2$ defined in the previous section.

\subsection{Comparison with the variational method}

Short-range non-relativistic interactions can often be approximated by an effective potential of finite range, for which a characteristic relation between the binding energy and the wave function at the origin is expected. The variational method provides a simple way to estimate this relation, with a mild dependence on the assumed trial wave function.

For a finite-range attractive potential we first adopt a simple exponential trial wave function $\Psi (r) =  c  e^{- r/r_0}$, a variational parameter and $c$ is fixed by normalization. In
this case one finds that for the expectation value of the non-relativistic Hamiltonian operator (binding energy) %
\begin{equation}
  E_B = \left|  \langle \Psi |\hat H_{\rm nonrel.} |\Psi \rangle \right|\simeq   \left|-  \langle \Psi |\hat T |\Psi \rangle \right| \simeq    {1\over  m_\pi r_0^2}\,, 
\end{equation}
where $\hat H_{\rm nonrel.}$ is the sum of the potential operator $\hat V$ and the  kinetic energy operator $\hat T$.
In the last step, we have used  the  virial theorem for an attractive potential $2 \langle  \hat T  \rangle  = \langle  \vec r \cdot \nabla \hat V  \rangle \simeq   \langle  \hat V  \rangle $,  which
implies
$E_B \simeq \langle\hat T\rangle$.
 Solving for $r_0$ in  terms of the binding energy and inserting into $\Psi(0)$ yields
\begin{equation}
 \frac{|\Psi(0)|}{m_\pi^{3/2}}  \sim { 2\kappa^{3/4} \over \sqrt{4\pi} } \,,
\end{equation}
where we have again adopted the dimensionless binding parameter
$\kappa$. The numerical results are shown as the dashed line in right panel of Fig.~\ref{fig:poleLoc}, deviating from the those  obtained from the $T$-matrix method (solid black line) by a factor of a few. This is understandable since a wavefunction of this form, commonly used for the Coulomb potential, is not well-suited for finite-range interactions. 

For a shallow bound state the asymptotic behavior of the exact wave function at large distances becomes $e^{i k_B r}/r$ with imaginary binding momentum $k_B \equiv \sqrt{-2m E_B} $, i.e., an exponentially decaying tail proportional to $ e^{-|k_B|r}/r$ determined by the binding momentum. This tail carries essentially no information about $|\Psi(0)|$, so a trial function chosen to reproduce only the asymptotic form does not give a reliable estimate of the wave function at the origin either, as 
illustrated by the dotted line in the right panel of Fig.~\ref{fig:poleLoc}.
Interestingly, a variational ansatz that combines an algebraic core with an exponential tail  up to a pre-factor of wave function normalization
\begin{equation}
    \Psi (r) \propto {1 \over r+ m_\pi^{-1}}  + c\cdot \exp({-{m_\pi r \over 2} })\,,
\end{equation}
which agrees better with the T-matrix-based result for $|\Psi(0)|$. In our fits this typically requires $c \gg 1$, indicating that the wave function is nearly flat inside a sphere of radius $r \lesssim 2\,m_\pi^{-1}$, with the exponential fall-off setting in only outside this core region.

The results for $|\Psi(0)|$ as a function of the binding energy are   summarized in right panel of Fig.~\ref{fig:poleLoc}, including various trial functions. For shallow bound states ($\kappa \ll 1$) the T-matrix method and the variational estimate agree within an overall factor of order unity. For binding energies of order the freeze-out temperature, $E_B\sim T_{\rm fo}\sim m_\pi/20$, we obtain
$|\Psi(0)| \sim \mathcal{O}(0.1)\,m_\pi^{3/2}$, in good agreement with the values assumed in our previous analysis~\cite{Chu:2024rrv}.

\subsection{Implications  for DM phenomenology}

In this subsection, we connect  our findings to the particle phenomenology, focusing on the contribution of an $S$-wave pole in the $\pi \pi$ scattering and comparing it with the astrophysical bound on DM self-scattering.
What one actually gains from the existence of a $S$-wave pole  is a non-perturbative contribution to the amplitude of $\langle R_I | \hat {\mathcal T} |R_I\rangle$ following Eq.~\eqref{eq:TonRI}, with its tree-level expression given by Eq.~\eqref{eq:swaveLOM}. Because of concrete form of the two-pion state $|R_I\rangle$, the presence of such intermediate pole  only contributes to self-scattering processes with pair-wise equal flavors, $\pi^a \pi^a \to \pi^b\pi^b$ (including the channel $a=b$). The amplitudes are equal for any incoming/outgoing flavor $a,\, b =1,2,3,4,5$. 
We may, therefore,  consider concretely $\pi^1 \pi^1 \to \pi^2\pi^2$, whose scattering amplitude via the intermediate pole, can be expressed as 
\begin{equation}\label{eq:relationM}
  \langle \pi^2  | \langle \pi^2   |R_I\rangle  \,  \langle R_I | \hat {\mathcal T}  |R_I\rangle   \, \langle R_I   |\pi^1 \rangle  |\pi^1 \rangle      =   {1\over 5}  \langle R_I | \hat {\mathcal T}  |R_I\rangle \,. 
\end{equation}
The strength of this matrix element is in turn decided by the scattering amplitude of the flavor singlet channel, ${\mathcal M}_I (s,t,u)$, using Eq.~\eqref{eq:TonRI}.

We now proceed to establish a connection between ${\mathcal M}_I (s,t,u)$ and  the bound state wave function at the origin, $|\Psi(0)|$. This can be done in the non-relativistic limit, where Eq.~\eqref{vLO-14} directly yields
\begin{equation}
    {\mathcal M}_I (s,t,u) \simeq 2 T(s)\,, 
\end{equation}
in which $T(s)$ is the $S$-wave projection of the corresponding amplitude, given by Eq.~\eqref{LS-CHUA}. To make connection to the bound state wave function, we take the non-relativistic form of the amplitude via $s =E^2$, 
\begin{eqnarray}
  T(E^2) \simeq  {(32\pi^3 m_\pi^2)}{ \tilde v  \over 1+  \tilde v  \tilde G(E) }\,,
\end{eqnarray}
where the relations $\tilde G(E) \simeq  32\pi^3 m_\pi^2\, G(s)$ and $\tilde G (E) \tilde v = G(s) v$ have been used. 
The intermediate pole has a mass smaller than $2m_\pi$ and cannot decay into two pions, so there is no imaginary part in the denominator for $E\le 2m_\pi$.
Given the existence of a below-threshold pole, following~\cite{Gamermann:2009uq}, we may also write this as
\begin{equation}
 T(E^2)=   { g_{X\pi\pi}^2 \over E - (2m_\pi -E_B) }\,. 
\end{equation} 
 The squared effective bound state-two pion coupling $g^2_{X\pi\pi}$  is defined as the residue of the two-pion projected $T$-matrix at the bound-state pole, treated as a constant for $S$-wave scattering,
\begin{eqnarray}
 g^2_{X \pi\pi} &=& \lim_{\tiny E\to (2m_\pi -E_B)} T(E^2){\big   ( } E - (2m_\pi -E_B) {\big )}\notag \\
 &=&  {32\pi^3 m_\pi^2} ~\lim_{E\to (2m_\pi -E_B)} { E - (2m_\pi -E_B)  \over \tilde v^{-1} + \tilde G(E) } \notag \\
& = & {32\pi^3 m_\pi^2} \left. \left( {d \tilde G(E) \over dE }\right)^{-1} \right |_{E =2m_\pi -E_B}\,,    
\end{eqnarray}
where in the last step we have taken the derivative with respect to~$E$ for both the numerator and denominator at the pole.  In the non-relativistic limit, $\tilde v$ is usually treated as energy-independent. Recall that $\Psi(0)$ is proportional to $({d \tilde G/ dE })^{-1/2}$ at the pole, one obtains $g_{X\pi\pi} \propto \Psi(0)$, being consistent with the result from the method using the non-relativistic limit of the Bethe-Salpeter equation~\cite{Petraki:2015hla}.

Now one can express the scattering amplitude taking the approximation $E \simeq  2m_\pi$, which yields
\begin{equation}
    {\mathcal M}_I (4m_\pi^2 , 0,0 ) \simeq {64\pi^3 m_\pi^2  \over 2m_\pi - (2m_\pi -E_B) } \left.\left( {d \tilde G(E) \over dE }\right)^{-1}\right|_{E = 2m_\pi -E_B} = - {512\pi^6 m_\pi^2  \over E_B } \left|{ \Psi(0)  \over \tilde G(2m_\pi -E_B) }  \right|^2  \,. 
\end{equation}
Note that a negative sign has been adopted for the last expression, as $d \tilde G/ dE$ is negative at the pole below  threshold.
After including the $1/5$ factor from Eq.~\eqref{eq:relationM} for the $\pi^1 \pi^1 \to \pi^2\pi^2$ scattering amplitude  and multiplying a symmetry factor $1/2$ for identical final state particles, the  corresponding scattering cross section  is given by 
\begin{equation}
   \sigma_{11\to 22} =    {1\over 128\pi m_\pi^2 }\, \left|{1\over 5}{\mathcal M}_I (4m_\pi^2 , 0,0 ) \right|^2 \simeq   {2048 \pi^{11} m_\pi^2\over 25  E_B^2 } \left|{ \Psi(0)  \over \tilde G(2m_\pi -E_B) }  \right|^4 \,.
\end{equation}
Regarding DM phenomenology, relevant observable quantities typically rely on the total number of self-scattering events per time and unit volume. For the $S$-wave pole-dominated scattering of interest here, an averaged cross section can thus be defined as
\begin{equation}
  \langle \sigma_\text{self} \rangle ~~\equiv  ~~  {1\over 2}{\sum_{a}\sum_{b}  (n_{\pi^a})^2  \over  (n_{\pi})^2 }~\sigma_{11\to 22}  =  {1\over 2} \, \sigma_{11\to 22}\,  ,
\end{equation}
where we sum up all possible initial states $a$ and final states $b$, of which each process contributes the same as $\sigma_{11\to 22}$ through the intermediate pole, and the pre-factor ${1/2}$ is the usual symmetric factor introduced to avoid double-counting due to identical initial states. In addition, here we consider the case where the particle number density is equipartitioned across all five flavors, each therefore  constituting one-fifth of the total dark pion density~$n_{\pi}= \sum_a n_{\pi^a}$.
At last, we express the averaged self-scattering cross section normalized to benchmark values: 
\begin{equation}
{\langle \sigma_\text{self}   \rangle  \over m_\pi} \simeq 1\,\text{cm}^2/\text{g} \left({0.1  \over \kappa }\right)^2 \left({0.5 \,\text{GeV} \over m_\pi }\right)^3 \left({|\Psi (0) | \over 0.14 m_\pi^{3/2} }\right)^4   \left({0.02  \over |G(s)|_{s = (2m_\pi -\kappa m_\pi)^2 } }\right)^4\,,
\end{equation}
for which $\tilde G(E)$ and $E_B$ have been replaced with dimensionless parameters $G(s)$ and $\kappa$ for clarity. That is, for sub-GeV DM of the framework investigated above, reasonable parameter choices yield DM self-scattering cross sections that are able to address astrophysical small-scale puzzles; see \cite{Tulin:2017ara, Knapen:2017xzo, Perivolaropoulos:2021jda, Lee:2020eap, Cirelli:2024ssz, Bozorgnia:2024pwk} for recent reviews.  Due to the presence of the pole, for $m_\pi/f_\pi \sim 5$ we generally obtain  values of $\langle \sigma_\text{self} \rangle $ larger than the NLO results of \cite{Hansen:2015yaa} by a factor of~$\mathcal{O}(50)$.
In this respect, it is also worth mentioning that self-scattering cross section induced by a bound state has non-trivial velocity-dependence, which is phenomenologically interesting. A detailed exploration of the bound-state affected SIMP phenomenology is left for future work.

\section{Discussion and conclusion}
\label{ch5}

In this work we have addressed a basic but so far largely open question in SIMP dark sectors based on confining $\mathrm{Sp}(N_c)$ dynamics with $N_f=2$ Dirac fermions: does a $\pi\pi$ bound state  exist in the regime of parameters relevant for dark-pion SIMP phenomenology?  Focusing on the  $\mathrm{SU}(4)/\mathrm{Sp}(4)$ coset, we have constructed the low-energy $\pi\pi$ scattering amplitude in the flavor-singlet channel using the chiral Lagrangian with an $\mathrm{SU}(4)$-invariant mass term, including LO and NLO contributions in the chiral expansion.  Employing the on-shell Lippmann-Schwinger equation in the chiral-unitary framework, we have studied the analytic structure of the $S$-wave  T-matrix in the complex $s$-plane and searched for poles below the two-pion threshold.

Within this setup, we find that an $S$-wave pole in the flavor-singlet channel emerges below $s=4m_\pi^2$ once the effective coupling $m_\pi/f_\pi$ exceeds a critical value of order three.  This value is obtained from the NLO-unitarized amplitude, and is  lower than what would be inferred from the LO kernel alone, which would suggest a bound state only for $m_\pi/f_\pi \gtrsim 4.2$, in a regime where the LO chiral expansion is no longer under perturbative control.  Using the size of the NLO contribution and an estimate of the NNLO effects (via Gaussian variations of the LECs at the level of $10^{-4}$–$10^{-6}$) as a proxy for the truncation uncertainty, we find that the combined LO+NLO singlet amplitude dominates over higher orders for $  m_\pi/f_\pi \lesssim 4\text{--}5$ when $ s \lesssim 4m_\pi^2$, and the bound-state pole for $m_\pi/f_\pi \gtrsim 3$ therefore lies within the region where the chiral-unitary treatment is still quantitatively meaningful.  The same analysis also shows that for smaller couplings, $m_\pi/f_\pi \lesssim 2.5$, the amplitude remains perturbative and no pole approaches the two-pion threshold from below.

The parameter region where the singlet bound state appears corresponds in the underlying confining theory to the regime where the (degenerate) dark-quark masses satisfy $m_q \sim \Lambda$, with $\Lambda$ the confinement scale.  In the chiral regime $m_q \ll \Lambda$ one has $m_\pi/f_\pi \propto \sqrt{m_q/\Lambda} \ll 1$, so that the chiral interactions are too weak to form a bound $\pi\pi$ state.  Our result therefore supports the picture that in the SIMP-motivated regime with moderately heavy pseudo-Goldstone bosons, where $m_\pi/f_\pi$ is of order a few, the lightest scalar beyond the dark pions is naturally interpreted as a $\pi\pi$ molecule in the most attractive $S$-wave flavor-singlet channel, in close analogy to the sigma meson in QCD when the physical pion mass is hypothetically increased into the few-hundred-MeV range.
We have also examined the $P$-wave amplitude in the non-singlet channels, where  we conclude that, within the parameter region where the chiral-unitary framework is under control, the $P$-wave interaction is too weak to support an $\ell=1$ bound state solely built from two dark pions.

To connect with the phenomenological role of a scalar $S$-wave $\pi\pi$ bound state in SIMP freeze-out, we have estimated the bound-state wave function at the origin, $|\Psi(0)|$, which controls bound-state-assisted decay and annihilation processes mediated by local operators. Using the non-relativistic reduction of the on-shell T-matrix near the pole, and the relation between the residue and the bound-state wave function derived in the separable-potential framework, we have obtained $|\Psi(0)|$.  This field-theoretic determination was then cross-checked against simple variational estimates.  For shallow binding, $\kappa \equiv E_B/m_\pi \ll 1$, the two methods agree within an overall factor of a few. 
In particular, for binding energies of order the typical SIMP freeze-out temperature,
$  E_B \sim T_{\rm fo}\sim m_\pi/20$, we find
$  |\Psi(0)| \;\sim\; \mathcal{O}(0.1)\,m_\pi^{3/2}$. This range is fully compatible with the parametric assumptions underlying the previous analysis of bound-state-catalyzed SIMP freeze-out~\cite{Chu:2024rrv}, where the enhancement of $4\to 2$ annihilation channels via intermediate $X=[\pi\pi]$ bound-states was encoded through $|\Psi(0)|$.
Beyond its role in number changing freeze-out, the same $S$-wave pole also induces a non-perturbative contribution to the late time dark pion self scattering rate. We showed that the corresponding average self-interaction cross section can be expressed in terms of the binding energy and the wave function at the origin. For the same class of shallow bound states that efficiently catalyze even numbered SIMP interactions at freeze-out, with $\kappa \sim 0.1$ and $|\Psi(0)| \sim 0.1\,m_\pi^{3/2}$ for sub-GeV $m_\pi$, we find that $\langle \sigma_\text{self} \rangle / m_\pi$ naturally lies in the astrophysically interesting range of order $\mathcal{O}(1)\,\text{cm}^2/\text{g}$, compatible with addressing small-scale structure puzzles.

The present analysis can be extended and refined in several directions.  On the theoretical side, a dedicated lattice determination of the $\pi\pi$ spectrum in $\mathrm{Sp}(4)$ with degenerate quarks along the lines of~\cite{Dengler:2023szi} will allow a more precise matching of the chiral-unitary description, sharpening the prediction for the critical $m_\pi/f_\pi$ and the binding energy and clarifying the molecular vs.~compact nature of the scalar. Recent lattice extractions of NLO LEC combinations in the SU(4)/Sp(4) effective theory with non-degenerate quark masses~\cite{Kolesova:2025ghl} provide a useful benchmark for the size of higher-order effects, although they are not directly tailored to the degenerate quark mass limit studied here; a systematic implementation of this input is left to future work.  More broadly, it would be interesting to generalize our approach to other gauge groups and flavor contents, to include vector resonances and coupled-channel effects, and to embed the bound state into a full finite-temperature Boltzmann analysis of SIMP-like cosmology.  We leave these refinements to future work. In summary, our findings lend theoretical support to phenomenological SIMP models in which a single light scalar resonance, interpreted as a $\pi\pi$ bound state, plays a distinguished role in determining the relic abundance and self-interactions of dark pions.

\medskip
\paragraph*{Acknowledgements.} 
We thank Teng Ma for useful discussions. This research has
received funding support from the NSRF via the Program Management Unit for Human Resources \& Institutional Development, Research and Innovation [grant number B50G670100]. X.C.~is funded by the National Natural Science Foundation of China (grant No.\,E4146602), and the Fundamental Research Funds for the Central Universities (grant No.\,E4EQ6605X2 and E5ER6601A2).  J.P.~is funded/co-funded by the European Union (ERC, NLO-DM, 101044443). This work was also supported by the Research Network Quantum Aspects of Spacetime (TURIS).

\appendix
\section{Basis of two-quark and four-quark states}
\label{app:grouptheory} 

Here we adopt the fundamental representation of $\mathrm{SU}(4)$ defined in Eqs.~(A12)–(A14) of Ref.~\cite{Kulkarni:2022bvh}. In this basis, among the 15 generators of $\mathrm{SU}(4)$,  the 10 unbroken generators in $\mathrm{Sp}(4)$ can be organized into several sets of $\mathrm{SU}(2)$ subalgebras:
\begin{equation}
    [\tilde T_u^1, \tilde T_u^2] = i \tilde T_u^3\,,\quad\text{and}\quad
    [\tilde T_d^1, \tilde T_d^2] = i \tilde T_d^3\,,
\end{equation}
and 
\begin{equation}
       [\tilde T^{11}, \tilde T^7] = i ({1\over 2} \tilde T_u^3 + {1\over 2} \tilde T_d^3) \,,\quad\text{and}\quad  [\tilde T^{13}, \tilde T^8] = i ({1\over 2} \tilde T_u^3 -  {1\over 2} \tilde T_d^3)\,. 
\end{equation}
This provides a Cartan–Weyl basis for the rank-2 group $\mathrm{Sp}(4)$, allowing the definition of weight vectors labeled by $(I_u, I_d)$, as well as the corresponding raising/lowering operators 
\begin{equation}
    T_u^\pm =\tilde T_u^1 \pm i \tilde T_u^2\,,~\text{~and~~}    T_d^\pm =\tilde T_d^1 \pm i \tilde T_d^2\,, 
\end{equation}
and the raising/lowering operators that change the third components of both $I_u$ and $I_d$ by one half, along either the diagonal (D) or off-diagonal (O) directions in the $(I_u, I_d)$ plane:
\begin{equation}
     T_D^\pm =\tilde T^{11} \pm i \tilde T^7\,,~\text{~and~~}    T_O^\pm =\tilde T^{13} \pm i \tilde T^8\,.
\end{equation}

\subsection{Single-quark states}

In the basis above, the vector of four Weyl fermions, which differs by an interchange of the second and third components relative to the convention used in the main text of Ref.~\cite{Kulkarni:2022bvh}, is written as
\begin{equation}
 \tilde \Psi = \begin{pmatrix}
u_L \\
\sigma^2 S u^*_R\\
d_L \\
\sigma^2 S d^*_R
\end{pmatrix}\,,
\end{equation}
where $S$ is the color matrix, as defined in Ref.~\cite{Kulkarni:2022bvh}.
The corresponding {\it isospin} assignments with respect to the two SU(2) subalgebras above, labeled as $(I_u, I^3_u; I_d, I^3_d)$, are $(+1/2,+1/2; 0, 0)$ for $u_L$, $(+1/2, -1/2; 0, 0)$ for $u^*_R$, $(0, 0;  +1/2, +1/2)$ for $d_L$, $(0, 0; +1/2, -1/2)$  for $d^*_R$, respectively. 
Adopting the fundamental representation of this basis on $\tilde \Psi$, the raising/lowering operators above change the four Weyl fermions as follows:
\begin{equation}
T_u^+ \begin{pmatrix}
u_L \\
u^*_R\\
d_L \\
 d^*_R  
\end{pmatrix} = \begin{pmatrix}
0\\
u^L\\
0 \\
0 
\end{pmatrix} \,,~~ T_u^- \begin{pmatrix}
u_L \\
u^*_R\\
d_L \\
 d^*_R  
\end{pmatrix} = \begin{pmatrix}
u^*_R\\
0\\
0 \\
0 
\end{pmatrix} \,,~~T_d^+ \begin{pmatrix}
u_L \\
u^*_R\\
d_L \\
 d^*_R  
\end{pmatrix} = \begin{pmatrix}
0\\
0\\
0 \\
d_L 
\end{pmatrix} \,,~~ T_d^- \begin{pmatrix}
u_L \\
u^*_R\\
d_L \\
 d^*_R  
\end{pmatrix} = \begin{pmatrix}
0\\
0\\
 d^*_R  \\
0 
\end{pmatrix} \,,
\end{equation}
and 
\begin{equation}
T_D^+ \begin{pmatrix}
u_L \\
u^*_R\\
d_L \\
 d^*_R  
\end{pmatrix} = {1\over \sqrt{2} } \begin{pmatrix}
d^*_R\\
0\\
u^*_R \\
0 
\end{pmatrix} \,,~ T_D^- \begin{pmatrix}
u_L \\
u^*_R\\
d_L \\
 d^*_R  
\end{pmatrix} = {1\over \sqrt{2} } \begin{pmatrix}
0\\
d_L\\
0 \\
u_L 
\end{pmatrix} \,,~T_O^+ \begin{pmatrix}
u_L \\
u^*_R\\
d_L \\
 d^*_R  
\end{pmatrix} ={1\over \sqrt{2} }  \begin{pmatrix}
d_L\\
0\\
0 \\
-u^*_R 
\end{pmatrix} \,,~ T_O^- \begin{pmatrix}
u_L \\
u^*_R\\
d_L \\
 d^*_R  
\end{pmatrix} ={1\over \sqrt{2} } \begin{pmatrix}
0\\
-d^*_R\\
u_L  \\
0 
\end{pmatrix} \,.
\end{equation}
The equations above directly illustrate that $T_D^+ | +1/2,+1/2; 0, 0 \rangle  =  {1\over \sqrt{2} }| 0, 0 ; +1/2, -1/2 \rangle  $, acting along the off-diagonal line,  and $T_O^+ | +1/2, +1/2 ; 0, 0 \rangle  =  {1\over \sqrt{2} } | 0, 0; +1/2,+1/2  \rangle  $ acting along the off-diagonal line in the ($I^3_u$, $I^3_d$)-plane.   Both conserve the sum $I \equiv  I_u + I_d$.  
In analogy with the SM isospin, the conjugates of the four Weyl fermions have the same isospin assignments and   transform in the same way, once they are arranged into the vector $(- u_R,\, u^*_L,\, -d_R,\, d^*_L)^T$.

\subsection{One-pion states}

Since our analysis of the two-pion sector singles out the flavor singlet combination, it is natural to already consider the singlets at the one-pion level, as it prepares the grounds for the subsequent state construction. The quark bilinears are organized into irreducible representations of the $\mathrm{SU}(2)_{u,d}$ subalgebra. This construction allows one to distinguish a genuine $\mathrm{SU}(2)_{u,d}$ singlet---which is annihilated by all raising and lowering operators---from neutral states that are simply the $I_u^3 =I_d^3 = 0$ components of nontrivial isospin multiplets.

\begin{figure}[t]
\includegraphics[width=8cm]{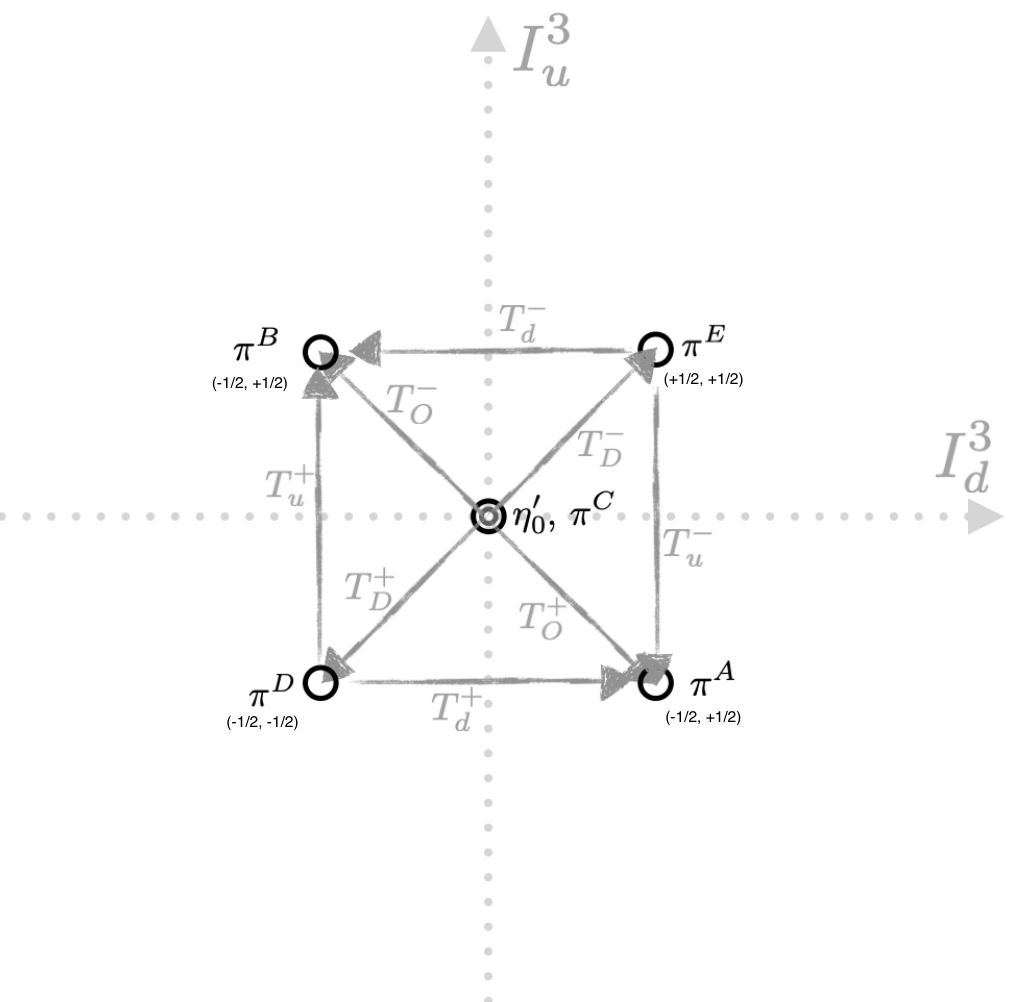}~~~~~
\includegraphics[width=8.1cm]{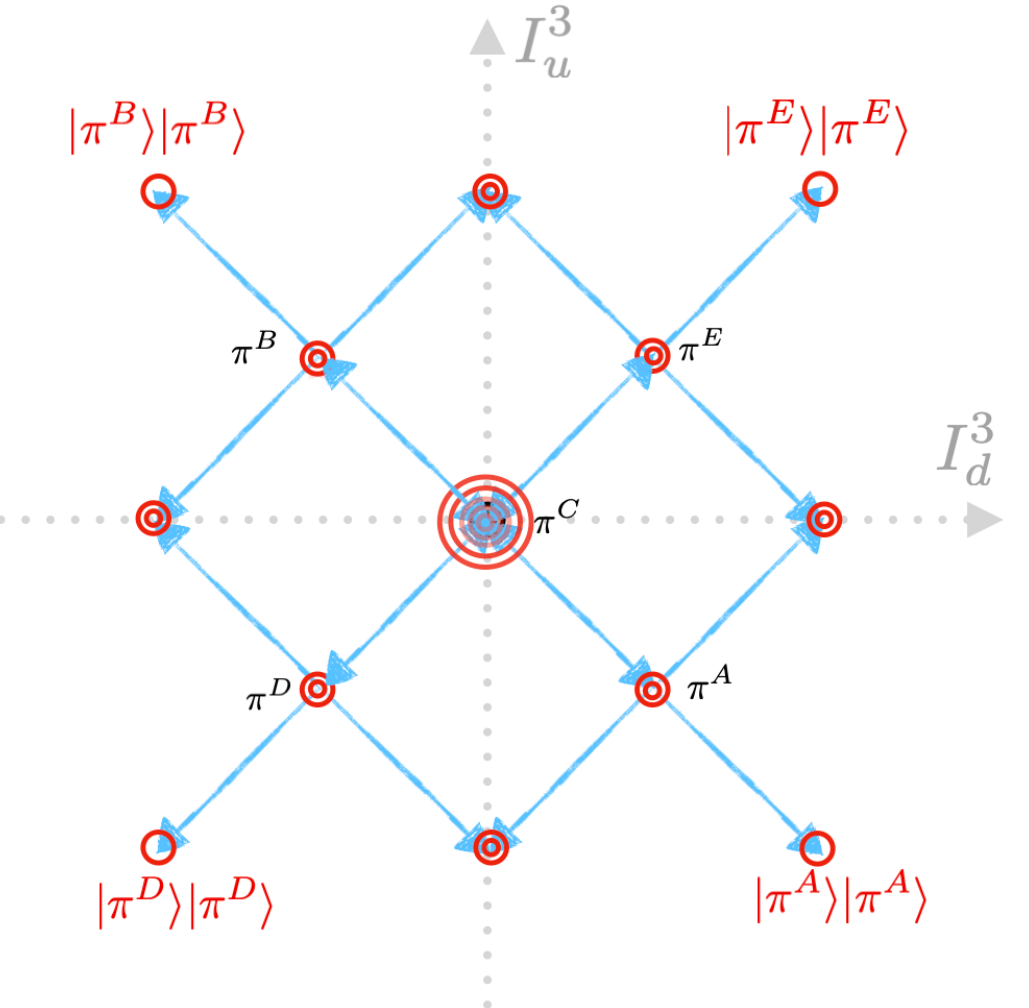}~~
\caption{Two-quark and four-quark states. {\it Left:} the singlet $\eta_0'$ and quintuplet $\pi^{A,\dots,E}$ of two-quark states (circles in black). {\it Right:} four-quark states  (circles in red) constructed from the two quintuplets: ${\bf 5}\; \otimes \; {\bf 5} = {\bf 1} \;\oplus {\bf 10} \;\oplus {\bf 14}$. Here each possible combination of two pion states (one circle of red) is depicted schematically by adding another pion (indicated by a blue arrow according to its weight vector), to a pre-existing one (labeled as $\pi^I$ in black). }
\label{Fig:isospin}
\end{figure}

For the two-quark states defined in Tab.~4 of Ref.~\cite{Kulkarni:2022bvh}, there exists the $\eta'_0$ state, $ \eta'_0 \equiv  {1\over \sqrt{2} }(\bar u \gamma_5 u + \bar d \gamma_5 d )$, which is invariant under all raising and lowering operators introduced above. It is therefore a genuine flavor singlet under $\mathrm{SU}(2)_{u,d}$, in direct analogy with the $\eta'$ in QCD, and is expected to receive a parametrically larger mass from the chiral anomaly. As a consequence, $\eta'_0$ does not belong to the light pseudo-Nambu-Goldstone multiplet associated with the $\mathrm{SU}(4)/\mathrm{Sp}(4)$ breaking pattern and it is natural to treat it as a heavier state that can be integrated out in the low energy description.
In contrast, the orthogonal combination  
\begin{equation}
    \pi^C \equiv {1\over  \sqrt{2} }(\bar u \gamma_5 u - \bar d \gamma_5 d ) = {1\over  \sqrt{2}  }(u_R^*  u_L - u_L^*  u_R  -  d_R^*   d_L + d_L^*  d_R) 
\end{equation}
 also has $I^3_u = I^3_d = 0$, but is not flavor-blind. It transforms non-trivially under the $\mathrm{SU}(2)_{u,d}$ sub-algebra and is the neutral component of an isospin multiplet. For instance, the action of the raising operator
 $T^+_O$ on $\pi^C$,
\begin{equation}
 T^+_ O    \pi^C =  {1\over  \sqrt{2}  }[ (T^+_ O    u_R^*)  u_L -  (T^+_ O   u_L^* ) u_R  -  (T^+_ O    d_R^*  ) d_L +  (T^+_ O   d_L^*)  d_R] +{1\over  \sqrt{2}  }[u_R^* (T^+_ O  u_L )- u_L^*  (T^+_ O u_R ) -  d_R^*  (T^+_ O  d_L) + d_L^* (T^+_ O  d_R)]\, \notag
\end{equation}
yields another pion state, $\pi^A \equiv \bar u \gamma_5 d$, where the $ T^+_ O$ operator acts non-trivially on the first (second) state in the first (second) square bracket. 
One  finds
\begin{equation}
 T^+_O    \pi^A =  0 \text{\,,~~}T^+_O    \pi^B =  - \pi^C \text{\,,~~} T^+_O    \pi^C =  \pi^A \text{\,,~~~~} T^-_O    \pi^A =  \pi^C \text{\,,~~} T^-_O    \pi^B = 0 \text{\,,~~and ~}  T^-_ O    \pi^C = - \pi^B\,,
\end{equation}
and  similar relations can be obtained for $T^\pm_D$.  For sub-algebra $\mathrm{SU}(2)_{u,d}$, we have
\begin{equation}
 T^+_d    \pi^{B} =  \pi^{E}\text{\,,~~~}  T^-_d   \pi^{E} =  \pi^{B}   \,,~~~ T^-_d    \pi^{A} =  - \pi^{D} \text{\,,~~and~~} T^+_d   \pi^{D} = - \pi^{A}\,,
\end{equation}
and, again, similar relations hold for  $T^\pm_u$. These relations are also demonstrated in the left panel of Fig.~\ref{Fig:isospin}.

\subsection{The 14-plet of two-pion states}

For four-quark states, it is easy to see that the states $|\pi^A\rangle |\pi^A\rangle $, $|\pi^B\rangle |\pi^B\rangle $, $|\pi^D\rangle |\pi^D\rangle $, and $|\pi^E\rangle |\pi^E\rangle $ have the highest weights in the $(I_u^3\,,\,I_d^3)$ plane, as illustrated in the right panel of Fig.~\ref{Fig:isospin}.  Other states are obtained by acting with lowering operators on them.  The resulting set of 14~linearly independent states is closed under the $\mathrm{Sp}(4)$ action, realizing the $\mathbf{14}$-dimensional representation, i.e., the mixed-symmetric channel $R_{MS}$.
Indeed, although not shown explicitly in the figure, $T^\pm_u$ and $T^\pm_d$ do not change the total $I$, and thus connect the symmetric states of the {\bf 14}-plet. For instance, from $|\pi^A\rangle \equiv |{1 \over 2}, -{1 \over 2}; {1 \over 2}, {1 \over 2}\rangle $, one obtains 
\begin{equation}
    T^-_d |\pi^A\rangle |\pi^A\rangle   =    T^+_d |1, -1; 1, 1\rangle = \sqrt{2}    |1, -1 ; 1, 0\rangle = -  |\pi^D\rangle |\pi^A\rangle -  |\pi^A\rangle |\pi^D \rangle \,. 
\end{equation}
So there are 
\begin{eqnarray}
|1, -1 ; 1, 0\rangle &=& - {1\over \sqrt{2}} \left( |\pi^D\rangle |\pi^A\rangle + |\pi^A\rangle |\pi^D \rangle \right) \,,\\
|1, 1 ; 1, 0\rangle &=& {1\over \sqrt{2} }\left( |\pi^B\rangle |\pi^E\rangle + |\pi^E\rangle |\pi^B \rangle \right)\,,\\
|1, 0 ; 1, -1\rangle &=& - {1\over \sqrt{2}} \left( |\pi^D\rangle |\pi^B\rangle + |\pi^B\rangle |\pi^D \rangle \right)\,,\\
|1, 0 ; 1, 1\rangle &=& {1\over \sqrt{2}} \left( |\pi^E\rangle |\pi^A\rangle + |\pi^A\rangle |\pi^E \rangle \right)\,,
\end{eqnarray}
where the overall negative sign does not play any physical role, and can be disregarded for simplicity.   Similarly,  there are states of the form $( |\pi^J \rangle |\pi^C \rangle + |\pi^C\rangle |\pi^J \rangle )/\sqrt{2}$ with $J = A, B, D, E$. The same eight states can also be derived from $ T^\pm_{D\,,O} |\pi^J \rangle |\pi^J \rangle $ for $J = A, B, D, E$, once the overall negative signs are omitted.

We now turn to the last two eigenstates of the {\bf 14}-plet, which have $I^3_u = I^3_d =0$. It is easier to obtain them with $ T^\pm_{D\,,O}$ as follows:
\begin{equation}\label{eq:14central1}
    T^-_O   ( |\pi^A \rangle |\pi^C \rangle + |\pi^C\rangle |\pi^A \rangle ) = |\pi^C \rangle |\pi^C \rangle - |\pi^A \rangle |\pi^B \rangle - |\pi^B \rangle |\pi^A \rangle + |\pi^C\rangle |\pi^C \rangle = - (  |\pi^A \rangle |\pi^B \rangle -2 |\pi^C\rangle |\pi^C \rangle + |\pi^B \rangle |\pi^A \rangle )\,,
\end{equation}
and, similarly,   
\begin{equation}\label{eq:14central2}
    T^-_D   ( |\pi^D \rangle |\pi^C \rangle + |\pi^C\rangle |\pi^D \rangle )=   |\pi^D \rangle |\pi^E \rangle -2 |\pi^C\rangle |\pi^C \rangle + |\pi^E \rangle |\pi^D\rangle  \,.
\end{equation}
In addition, if the global $\mathrm{Sp}(4)$  flavor symmetry  is exact, the basis can be arbitrarily rotated. It is therefore equivalent to choose, first, the sum of Eq.~\eqref{eq:14central1} and \eqref{eq:14central2} 
\begin{equation}~\label{eq:neutral1}
- |\pi^A\rangle |\pi^B\rangle -  |\pi^B\rangle |\pi^A \rangle  +  |\pi^D\rangle |\pi^E\rangle +  |\pi^E\rangle |\pi^D \rangle  \,,
\end{equation}
which coincides with $T^-_d | 1, 0 ; 1, 1\rangle$ up to a normalization factor, and, second, the difference
\begin{equation} ~\label{eq:neutral2}
 |\pi^A\rangle |\pi^B \rangle  + |\pi^B\rangle |\pi^A\rangle   - 4 |\pi^C\rangle |\pi^C \rangle  +  |\pi^D\rangle |\pi^E \rangle   + |\pi^E\rangle |\pi^D  \rangle   \,.
\end{equation}
Note that gauging quarks under a $U(1)$ symmetry and assigning different charges to $u$ and $d$  fixes the physical states; see, e.g., Tab.~2 of Ref.~\cite{Kulkarni:2022bvh}.

\subsection{The singlet and 10-plet of two-pion states}

The singlet in the two-pion states, via ${\bf 5}\otimes{\bf 5} = {\bf 1} \oplus {\bf 10} \oplus {\bf 14}$, must be symmetric under interchange of the two pions. In the neutral sector with $I_u^3 = I_d^3 = 0$ there are three symmetric states: two of them belong to the {\bf 14}-plet and have been constructed above, while the third should be  the singlet.
Imposing orthogonality to the {\bf 14}-plet states and normalizing it fixes the singlet uniquely. Direct computation yields 
\begin{equation}~\label{eq:neutral3}
{1 \over \sqrt{5} } \left(| \pi^A \rangle \pi^B \rangle +  | \pi^B \rangle \pi^A \rangle + |\pi^C \rangle |\pi^C\rangle +  | \pi^D \rangle \pi^E \rangle +  | \pi^E \rangle \pi^D \rangle \right) \,.
\end{equation}
This is in agreement
with Tab.~3 of \cite{Bijnens:2011fm} (see also Ref.~\cite{Drach:2021uhl}) which states the two-meson singlet as,
\begin{equation}
   |  R_I \rangle \equiv  \frac{\text{Tr}[T^a  T^b] }{\sqrt{(2\cdot 2+1)(2-1)} }  | \pi^a \rangle  |\pi^b \rangle = \sum_{a = 1}^5\frac{1 }{\sqrt{5} } |\pi^a \rangle | \pi^a \rangle  \,.
\end{equation}
We can rewrite this state in the $\{\pi^{A,B,C,D,E}\}$ basis. Using
\begin{equation}
    \pi^{A,B} = {1\over \sqrt{2}}\left( \pi^1 \pm i \pi^2 \right)\,, ~~\pi^C = \pi^3\,, ~~\pi^{D,E} = {1\over \sqrt{2}}\left( \pi^5 \pm i \pi^4 \right)\,.
\end{equation}
and the relations $\pi^A = (\pi^B)^\dagger$ and $\pi^D = (\pi^E)^\dagger$, one also finds that the states with  $I_u^3 = I_d^3 = 0$, such as the three given by Eqs.~(\ref{eq:neutral1}), (\ref{eq:neutral2}),  and (\ref{eq:neutral3}), are automatically self-conjugate, as expected.  
At last, the  {\bf 10}-plet states are antisymmetric combinations of two pions as 
\begin{equation}
  {1\over \sqrt{2} } \left( |\pi^I \rangle | \pi^J \rangle - |\pi^J \rangle | \pi^I \rangle \right) \,,
\end{equation}
where $I\neq J = A,\,B,\,C,\,D,\,E$. %

\section{Amplitudes and their \boldmath$S$-wave Projections}
\label{app:amplitudes}

As mentioned in the main text, we have adopted the expressions of these $\pi\pi$ scattering amplitudes at various orders from Ref.~\cite{Bijnens:2011fm}. For completeness and clarity, we provide here the details of the the amplitudes, as well as their $S$-wave projections. 

\subsection{Definition of useful variables and functions}

Before going to the expressions of the amplitudes, useful variables and functions have to be introduced, which  simplify the expressions in the following subsection. 
First, we assign the Mandelstam variables for two-body scattering of $\pi^{a_1} (p_1) + \pi^{a_2} (p_2) \to \pi^{a_3} (p_3)+ \pi^{a_4} (p_4) $ as
\begin{eqnarray}
s &=& (p_1 + p_2)^2 = (p_3 + p_4)^2 \equiv w^2\,,
\notag\\
t &=& (p_1 - p_3)^2 = (p_4 - p_2)^2 \,,
\notag\\
u &=& (p_1 - p_4)^2 = (p_3 - p_2)^2\,,
\label{mandel-1}
\end{eqnarray}
where a new variable $w$ for the  center-of-mass energy is also introduced. The angular dependence in the center-of-mass frame for two-body elastic scattering is, as usual, $t = -2\,{\bf p}^2\left( 1 - \cos\theta\right)$, and $u = -2\,{\bf p}^2\left( 1 + \cos\theta\right)$, 
where  ${\bf p}$  and $\theta$ are the incoming 3-momentum and scattering angle in the center-of-mass frame, and 
\begin{eqnarray}
{\bf p}^2 =\frac14\left( s - 4m_\pi^2\right)=\frac{1}{4}\,s\,\sigma^2(s)\,,\qquad \sigma(s)=  i \tilde \sigma(s)= \sqrt{ 1- \frac{4\,m_\pi^2}{s}  }\,.
\end{eqnarray}
The  functions  $\tilde \sigma(s)$ and $\sigma(s)$ will be used for the kinematic regions of $s$ below and above two-pion threshold, respectively.

Several other useful functions are defined as
\begin{eqnarray}
\bar{I}(s) ~\equiv  ~  \sigma(s)  \ln \left[\frac{\sigma(s) -1}{\sigma(s) +1}\right]\,, \quad ~\text{and}~~\bar{J}(s) ~\equiv  ~  \frac{1}{16 \pi^2}\left[\bar{I}(s) +2\right] \,.
\end{eqnarray}
The exact calculation of the Chew-Mandelstam function, $\bar{J}(s)$, depends on the value of $s$, whether $0<s< 4m_\pi^2$ or $s > 4m_\pi^2$. So we correspondingly adopt the notation that 
\begin{eqnarray}\label{eq:Jintegral}
{\bar J}(s) ~\equiv~
\begin{cases}
{\bar J}_{-}(s) & \text{for}\quad 0<s<4m_\pi^2\,,\\
{\bar J}_{+}(s) & \text{for}\quad s>4m_\pi^2\,.
\end{cases}
\end{eqnarray}
For the corresponding ranges of $s$, one obtains~\cite{Knecht:1995tr} 
\begin{equation}
\bar{J}_-(s)  = \frac{1}{8\pi^2} \left[ 1 - \tilde \sigma(s) \tan^{-1}\left( {1\over \tilde \sigma(s) } \right) \right]\,
\end{equation}
and 
\begin{equation}
\bar{J}_+(s)   =   \frac{1}{16\pi^2} {\Big[} 2 + \sigma(s) {\rm Ln}(s)  + i \pi \sigma(s) {\Big]}\, ,
\end{equation}
where, for $s\ge 4m_\pi^2$, we have defined 
\begin{equation}
    {\rm Ln}(s) ~\equiv  ~ \ln\left(\frac{1-\sigma(s)}{1 + \sigma(s)}\right)\,.
\end{equation}

Furthermore, in order to project the $S$-wave components from the one-loop amplitudes, one always encounters the integration over $\cos\theta$ involving the loop-functions, $\bar{J}(t)$ and $\bar{J}(u)$, while integrating $\bar{J}(s)$ over $\cos\theta$  is trivial and will not be further explained. 
Thus, before ending this subsection, we will perform the integration of the loop-function $\bar{J}(t)$, to express them eventually as functions of $s$. The results are the same for  $\bar{J}(u)$, by introducing the notation $\mathcal{J}_{0,1,2}(s)$ from 
\begin{eqnarray}\label{eq:Jintegral}
\mathcal{J}_{i}(s) &\equiv &  \int_{-1}^{1}d(\cos\theta)\,t^i \bar{J} (t) \equiv 
\begin{cases}
\mathcal{J}_{i-}(s) & \text{for}\quad 0<s<4m_\pi^2\,,\\
\mathcal{J}_{i+}(s) & \text{for}\quad s>4m_\pi^2\,.
\end{cases}
\end{eqnarray}
Below the two-pion threshold, one finds
\begin{eqnarray}
\mathcal{J}_{0-}(s) &\equiv& \int_{-1}^{1}d(\cos\theta)\,\bar{J}(t) = \frac{1}{8 \pi ^2 s \tilde{\sigma }^2}
\left[3 s\tilde{\sigma }^2 -4 m_{\pi }^2 \tan ^{-1}\left(\tilde{\sigma }\right)^2 - 2 s \tilde{\sigma } \tan ^{-1}\left(\tilde{\sigma
   }\right)\right]\,,
\nonumber\\
\mathcal{J}_{1-}(s) &\equiv &\int_{-1}^{1}d(\cos\theta)\,t\,\bar{J}(t) = \frac{1}{32 \pi ^2 s \tilde{\sigma }^2}\Big[ \tilde{\sigma } \left(16 m_{\pi }^4 \left(4 - \tan
   ^{-1}\left(\tilde{\sigma }\right)^2 \right)-36 m_{\pi }^2 s+5
   s^2\right)
\nonumber\\
&&\qquad\qquad\qquad\qquad\qquad\qquad\qquad -\,4 \left(8 m_{\pi }^4 -6 m_{\pi }^2 s + s^2\right) \tan
   ^{-1}\left(\tilde{\sigma }\right) \Big]\,,
\nonumber\\
\mathcal{J}_{2-}(s) &=& \int_{-1}^{1}d(\cos\theta)\,t^2\,\bar{J}(t) = \frac{1}{144 \pi ^2 s \tilde{\sigma }^2}
\Big[ 12 s \tilde{\sigma } \left(7 m_{\pi }^2 s - 6 m_{\pi
   }^4 - s^2\right) \tan ^{-1}\left(\tilde{\sigma }\right) - 16 m_{\pi }^6
   \left(9 \tan ^{-1}\left(\tilde{\sigma }\right)^2-44\right)\,,
\nonumber\\
&&\qquad\qquad\qquad\qquad\qquad\qquad\qquad  +\, 165 m_{\pi
   }^2 s^2 - 612 m_{\pi }^4 s - 14 s^3\Big]\,,
\nonumber\\
\mathcal{J}_{3-}(s) &\equiv& \int_{-1}^{1}d(\cos\theta)\,t^3\,\bar{J}(t) = \frac{1}{1152 \pi ^2 s\tilde\sigma}\Big[ 16384 m^8-19152 m^6 s+7620 m^4 s^2-1288 m^2 s^3
\nonumber\\
&&\qquad\qquad\qquad\qquad\qquad\qquad\qquad -\,2880 m^8 \tan
   ^{-1}\left(\tilde\sigma\right)^2
\nonumber\\
&&\qquad\qquad\qquad\qquad\qquad\qquad\qquad -\,24 s \tilde\sigma
   \left(60 m^6-118 m^4 s+34 m^2 s^2-3 s^3\right) \tan ^{-1}\left(\tilde\sigma\right)+81 s^4\Big]\,.\nonumber
\end{eqnarray}
Above the two-pion threshold, the $\mathcal{J}_{i+}$ functions read 
\begin{eqnarray}
\mathcal{J}_{0+}(s) &\equiv& \int_{-1}^1 d(\cos\theta)\,\bar{J}(t) = \pi_{16} \int_{-1}^1 d(\cos\theta)\left(\bar{I}(t) + 2 \right) = \pi_{16}\left(\mathcal{I}_0(s) + 4 \right)\,,
\nonumber\\
\mathcal{J}_{1+}(s) &\equiv& \int_{-1}^1 d(\cos\theta)\,t\,\bar{J}(t) = \pi_{16} \int_{-1}^1 d(\cos\theta)\left( t\,\bar{I}(t) + 2\,t \right) = \pi_{16}\left(\mathcal{I}_1(s) - 2\,s\,\sigma^2(s) \right)\,,
\nonumber\\
\mathcal{J}_{2+}(s) &\equiv& \int_{-1}^1 d(\cos\theta)\,t^2\,\bar{J}(t) = \pi_{16} \int_{-1}^1 d(\cos\theta)\left( t^2\,\bar{I}(t) + 2\,t^2 \right) = \pi_{16}\left(\mathcal{I}_2(s) + \frac43\,s^2\,\sigma^4(s) \right)\,,
\nonumber\\
\mathcal{J}_{3+}(s) &\equiv& \int_{-1}^1 d(\cos\theta)\,t^3\,\bar{J}(t) = \pi_{16} \int_{-1}^1 d(\cos\theta)\left( t^3\,\bar{I}(t) + 2\,t^3 \right) = \pi_{16}\left(\mathcal{I}_2(s) - s^3\,\sigma^6(s) \right)\,,\nonumber
\end{eqnarray}
where we have also used the integrals of the $\bar I$-function introduced above: 
\begin{eqnarray}
\mathcal{I}_0(s) &=& \int_{-1}^1 d(\cos\theta)\,\bar{I}(t)
\nonumber\\
&=& \frac{-2\,m_\pi^2}{s\,\sigma(s)\left(\sigma^2(s)-1\right)}\left[4\,\sigma^2(s) + 4\,\sigma(s)\,{\rm Ln}(s) +\left(\sigma^2(s) - 1\right)\,{\rm Ln}^2(s)\right]\,,
\nonumber\\
\mathcal{I}_1(s) &=& \int_{-1}^1 d(\cos\theta)\,t\,\bar{I}(t) 
\nonumber\\
&=&\frac{-2\,m_\pi^4}{s\,\sigma(s)\left(\sigma^2(s) -1 \right)^2}\left[4\,\sigma^2(s) + 4\,\sigma(s)\left(\sigma^2(s) + 1\right) {\rm Ln}(s)  
+ \left(\sigma^2(s) -1\right)^2\,{\rm Ln}^2(s)\right]\,,
\nonumber\\
\mathcal{I}_2(s) &=& \int_{-1}^1 d(\cos\theta)\,t^2\,\bar{I}(t) 
\nonumber\\
&=&\frac{-4\,m_\pi^6}{9\,s\,\sigma(s)\left( \sigma^2(s) -1 \right)^3}
\nonumber\\
&\times&\left[-4\,\sigma^2(s) \left(4\,\sigma^4(s) -21\,\sigma^2(s) + 9\right) + 12\,\sigma(s)\left(3\,\sigma^4(s) + 8\,\sigma^2(s) -3\right){\rm Ln}(s) + 9 \left(\sigma^2(s)-1\right)^3\,{\rm Ln}^2(s) \right]\,,
\nonumber\\
\mathcal{I}_3(s) &=& \int_{-1}^1 d(\cos\theta)\,t^3\,\bar I(t)
\nonumber\\
&=& \frac{-8 m_{\pi }^8}{9 s \sigma 
   \left(\sigma ^2-1\right)^4}
   \Big[-32 \sigma ^8+173 \sigma ^6-150 \sigma ^4+45 \sigma ^2+45
   \left(\sigma ^2-1\right)^4 \tanh ^{-1}(\sigma )^2
\nonumber\\
&&\qquad\qquad\qquad\qquad -\,6 \left(15 \sigma ^6+73 \sigma
   ^4-55 \sigma ^2+15\right) \sigma  \tanh ^{-1}(\sigma )\Big]\,.\nonumber
\end{eqnarray}

\subsection{Loop function of two-body propagators}\label{app-loop-function}

For two-body scattering with initial total four-momentum $P$, the loop function~$G(s)$ with masses $M_1$ and $M_2$ has the generic form~\cite{Oller:1998hw}
\begin{align}
G(s)
&=  i \int \frac{d^4 q}{(2 \pi)^4} \frac{1}{q^2-M_{1}^2+i \epsilon} \frac{1}{(P-q)^2-M_{2}^2+i \epsilon} \nonumber \\
&=  \frac{1}{32 \pi^2}\left[-\frac{\Delta}{s} \log \frac{M_1^2}{M_2^2}+\frac{\nu}{s}\left\{\log \frac{s-\Delta+\nu \sqrt{1+\frac{M_1^2}{q_{\text {max }}}}}{-s+\Delta+\nu \sqrt{1+\frac{M_1^2}{q_{\text {max }}}}}+\log \frac{s+\Delta+\nu \sqrt{1+\frac{M_2^2}{q_{\text {max }}^2}}}{-s-\Delta+\nu \sqrt{1+\frac{M_2^2}{q_{\text {max }}^2}}}\right\}\right. \nonumber\\
&\qquad +2 \frac{\Delta}{s} \log \frac{1+\sqrt{1+\frac{M_2^2}{q_{\text {max }}^2}}}{1+\sqrt{1+\frac{M_2^2}{q_{\text {max }}^2}}}-2 \log \left[\left(1+\sqrt{1+\frac{M_1^2}{q_{\text {max }}^2}}\right)\left(1+\sqrt{1+\frac{M_2^2}{q_{\text {max }}^2}}\right)\right]  \left.+\log \frac{M_1^2 M_2^2}{q_{\text {max }}^4}\,\right]\,,
\label{loop-function}
\end{align}
where $\nu= \left(s-\left(M_1+M_2\right)^2\right)^{1 \over  2}\left(s-\left(M_1-M_2\right)^2\right)^{1 \over 2}$ and $\Delta=M_2^2-M_1^2$. In the case of equal masses, $M_1=M_2=m_\pi$, the above formula reduces to
\begin{eqnarray}
G(s)=\frac{1}{(4 \pi)^2}\left[\sigma(s) \log \frac{\sigma(s) \sqrt{1+\frac{m_\pi^2}{q_{\max }^2}}+1}{\sigma(s) \sqrt{1+\frac{m_\pi^2}{q_{\text {max }}^2}}-1}-2 \log \left\{\frac{q_{\max }}{m_\pi}\left(1+\sqrt{1+\frac{m_\pi^2}{q_{\text {max }}^2}}\right)\right\}\,\right]\,,
\label{loop-function-2}
\end{eqnarray}
where $q_{\rm max}$ is the cut-off momentum, considered to be a free parameter that indicates the energy-scale of the system. In this work, we adopt $q_{\rm max} = 3m_\pi$ for the  numerical results.

\subsection{\boldmath$S$-wave projection of the scattering amplitude in the flavor singlet channel}
\label{s-wave-amp}
Below we proceed to compute the $S$-wave projection of the scattering amplitudes in the flavor singlet channel needed to evaluate the pole position of the two-pion bound state. 
Following Ref.~\cite{Oller:2000ma}, one obtains 
\begin{eqnarray}
T(s) &=& \textcolor{black}{\frac{1}{4}}\int_{-1}^{+1} d(\cos\theta)\,P_0(\cos\theta)\mathcal{M}_I(s, t, u) \,,
\nonumber\\
&=& \textcolor{black}{\frac{1}{4}}\int_{-1}^{+1} d(\cos\theta)\,\Big[\frac52\,\Big[ \mathcal{B}(s,t,u) + \mathcal{B}(t,u,s) - \frac{3}{5}\,\mathcal{B}(u,s,t) \Big] + 5\,\mathcal{C}(s,t,u) + \mathcal{C}(t,u,s) + \mathcal{C}(u,s,t)\, \Big]\,,
\label{int-kernel}
\end{eqnarray}
where $P_0(\cos\theta) =1$, and the functions $\mathcal{B}$ and $\mathcal{C}$ are defined by
\begin{eqnarray}
\mathcal{B}(s, t, u) &=& \mathcal{B}_{\LO}(s, t, u) + \mathcal{B}_{\NLO}(s, t, u) + \mathcal{B}_\text{NNLO}(s, t, u)\,,
\\
\mathcal{C}(s, t, u) &=& \mathcal{C}_{\LO}(s, t, u) + \mathcal{C}_{\NLO}(s, t, u)+ \mathcal{C}_\text{NNLO}(s, t, u)\,.
\label{def-B-C-s-wave}
\end{eqnarray}

Using the results of Ref.~\cite{Bijnens:2011fm}, one obtains for the LO amplitudes,
\begin{eqnarray}
\mathcal{B}_{\LO}(s, t, u) &=& \frac{1}{f_\pi^2}\left( -\frac12\,t + m_\pi^2 \right)\,,
\\
\mathcal{C}_{\LO}(s, t, u) &=& 0\,,
\label{def-B-C-LO}
\end{eqnarray}
and for the NLO amplitudes,
\begin{equation}
\begin{aligned}
& \mathcal{B}_{\NLO}(s, t, u)= \frac{1}{f_\pi^4}\left[\mathcal{B}_P(s, t, u) + \mathcal{B}_S(s, t-u) + \mathcal{B}_S(u, t-s) + \mathcal{B}_T(t)\right] \,, \\
& \mathcal{C}_{\NLO}(s, t, u)= \frac{1}{f_\pi^4}\left[\mathcal{C}_P(s, t, u) + \mathcal{C}_S(s) + \mathcal{C}_T(t) + \mathcal{C}_T(u)\right]\, .
\end{aligned}
\label{def-B-C-NLO}
\end{equation}
Here $\mathcal{B}_P(s, t, u)$ and $\mathcal{C}_P(s, t, u)$ are the polynomial parts, the remaining terms are collectively known as the unitarity correction. Following Ref.~\cite{Bijnens:2011fm}, we rewrite these parts in their simplest form, satisfying the symmetry constraints, 
\begin{equation}
\begin{aligned}
& \mathcal{B}_P(s, t, u)=\alpha_1\,m_\pi^4 +\alpha_2\,m_\pi^2\, t+\alpha_3 t^2+\alpha_4(s-u)^2 \,, \\
& \mathcal{C}_P(s, t, u)=\beta_1\,m_\pi^4 +\beta_2\,m_\pi^2\, s+\beta_3 s^2+\beta_4(t-u)^2 \,,
\end{aligned}
\label{def-BP-CP}
\end{equation}
in which the coefficients $\alpha_i$ and $\beta_i$ are given by 
\begin{eqnarray}
\alpha_1 &=& 16\left(L_0^r+L_8^r\right)+\left(\frac{7}{6}+\frac{1}{n}-\frac{2 n}{3}\right) L+\left(\frac{19}{18}-\frac{5 n}{9}+\frac{1}{n}\right) \pi_{16} \,,
\\
\alpha_2 &=& -16 L_0^r-4 L_5^r+\left(\frac{5 n}{12}-\frac{2}{3}\right) L\, \textcolor{black}{-}\left(\frac{5}{9}-\frac{11 n}{36}\right) \pi_{16} \,,
\\
\alpha_3 &=& 4 L_0^r+L_3^r+\left(\frac{1}{8}-\frac{n}{16}\right) L+\left(\frac{5}{48}-\frac{n}{24}\right) \pi_{16}  \,,
\\
\alpha_4 &=& L_3^r-\left(\frac{1}{24}+\frac{n}{48}\right) L-\left(\frac{5}{144}+\frac{n}{36}\right) \pi_{16}\,,
\\
\beta_1 &=& 32\left(L_1^r-L_4^r+L_6^r\right)-\frac{1}{2 n^2}\left(L+\pi_{16}\right),
\\
\beta_2 &=& 16\left(L_4^r-2 L_1^r\right) ,
\\
\beta_3 &=& 8 L_1^r+2 L_2^r-\frac{3}{16} \pi_{16}-\frac{3}{16} L \,,
\\
\beta_4 &=& 2 L_2^r-\frac{1}{16} L-\frac{1}{16} \pi_{16}\,,
\end{eqnarray}
where we have introduced the following short-hand notation for the logarithmic terms arising from the loop diagrams with respect to the dimensional regularization energy scale $\mu$ as 
\begin{equation}
L=\pi_{16} \log \left(\frac{m_\pi^2}{\mu^2}\right), \quad \pi_{16}=\frac{1}{16 \pi^2} .
\end{equation}
Above, we have fixed a typo of relative sign in front of the $\pi_{16}$ from Ref.\,\cite{Bijnens:2011fm} for the $\alpha_2$ term, as pointed out by  Ref.\,\cite{Hansen:2015yaa}. The remaining terms, $\mathcal{B}_S$, $\mathcal{B}_T$, $\mathcal{C}_S$, and $\mathcal{C}_T$ are written as follows: 
\begin{eqnarray}
\mathcal{B}_S(s,t-u) &=& \bar{J}(s)\left[-\frac{1}{2 n}\,m_\pi^4 -\frac{1}{4}\,m_\pi^2\, s + \frac{1}{16}(n+1)\,s^2 +\frac{1}{12}(n-1)\left(m_\pi^2-\frac{s}{4}\right)(t-u)\right],
\\
\mathcal{B}_S(u,t-s) &=& \bar{J}(u)\left[-\frac{1}{2 n}\,m_\pi^4 -\frac{1}{4}\,m_\pi^2\, u + \frac{1}{16}(n+1)\,u^2 +\frac{1}{12}(n-1)\left(m_\pi^2-\frac{u}{4}\right)(t-s)\right], 
\\
\mathcal{B}_S(t,u-s) &=& \bar{J}(t)\left[-\frac{1}{2 n}\,m_\pi^4 -\frac{1}{4}\,m_\pi^2\, t + \frac{1}{16}(n+1)\,t^2 +\frac{1}{12}(n-1)\left(m_\pi^2-\frac{t}{4}\right)(u-s)\right],
\\
\mathcal{B}_S(s,u-t) &=& \bar{J}(s)\left[-\frac{1}{2 n}\,m_\pi^4 -\frac{1}{4}\,m_\pi^2\, s + \frac{1}{16}(n+1)\,s^2 +\frac{1}{12}(n-1)\left(m_\pi^2-\frac{s}{4}\right)(u-t)\right],  
\\
\mathcal{B}_S(u,s-t) &=& \bar{J}(u)\left[-\frac{1}{2 n}\,m_\pi^4 -\frac{1}{4}\,m_\pi^2\, u + \frac{1}{16}(n+1)\,u^2 +\frac{1}{12}(n-1)\left(m_\pi^2-\frac{u}{4}\right)(s-t)\right],
\\
\mathcal{B}_S(t,s-u) &=& \bar{J}(t)\left[-\frac{1}{2 n}\,m_\pi^4 -\frac{1}{4}\,m_\pi^2\, t + \frac{1}{16}(n+1)\,t^2 +\frac{1}{12}(n-1)\left(m_\pi^2-\frac{t}{4}\right)(s-u)\right], 
\end{eqnarray}
and 
\begin{eqnarray}
\mathcal{B}_T(t) &=& -\frac{1}{8} \bar{J}(t)(t-2\,m_\pi^2)^2\,, 
\\
\mathcal{C}_S(s) &=& \bar{J}(s)\left(\frac{1}{2 n^2}\,m_\pi^4 +\frac{1}{8} s^2\right), 
\\
\mathcal{C}_T(t) &=& \frac{1}{8} \bar{J}(t)(t-2\,m_\pi^2)^2 \,,
\\
\mathcal{C}_T(u) &=& \frac{1}{8} \bar{J}(t)(u-2\,m_\pi^2)^2\,.
\end{eqnarray}
Here, the value of $n$ is determined by the symmetry breaking pattern of $\text{SU}(2n) \to \text{Sp}(2n)$. We  adopt $n=2$ below, corresponding to $N_f =2$ of the main text, to derive the final expressions of amplitudes for the concrete model of the work. Finally, we note that  the analytical expressions at NNLO are significantly more lengthy and will thus not be reproduced here. We direct the interested reader to Sec.\,4.3 and App.\,D of Ref.~\cite{Bijnens:2011fm}.  
\subsubsection{$S$-wave projection of scattering amplitude from tree-level diagrams}
At LO, we obtain the following $S$-wave projected expressions as
\allowdisplaybreaks
\begin{eqnarray}
\mathcal{B}_{\LO}^{(s)}(s) &\equiv& \textcolor{black}{\frac{1}{4}}\int_{-1}^{+1} d(\cos\theta)\,\mathcal{B}_{\LO}(s, t, u) = \textcolor{black}{\frac{1}{4}}\int_{-1}^{+1} d(\cos\theta)\,\frac{1}{f_\pi^2}\left( -\frac12\,t + m_\pi^2 \right)
= \frac{1}{\textcolor{black}{4}\,f_\pi^2}\left( \frac12\,s\,\sigma^2 + m_\pi^2 \right),
\nonumber\\
\mathcal{B}_{\LO}^{(t)}(s) &\equiv& \textcolor{black}{\frac{1}{4}}\int_{-1}^{+1} d(\cos\theta)\,\mathcal{B}_{\LO}(t, u, s) = \textcolor{black}{\frac{1}{4}}\int_{-1}^{+1} d(\cos\theta)\,\frac{1}{f_\pi^2}\left( -\frac12\,u + m_\pi^2 \right)
= \frac{1}{\textcolor{black}{4}\,f_\pi^2}\left( \frac12\,s\,\sigma^2 + m_\pi^2 \right),
\nonumber\\
\mathcal{B}_{\LO}^{(u)}(s) &\equiv& \textcolor{black}{\frac{1}{4}}\int_{-1}^{+1} d(\cos\theta)\,\mathcal{B}_{\LO}(u, s, t) = \textcolor{black}{\frac{1}{4}}\int d(\cos\theta)\,\frac{1}{f_\pi^2}\left( -\frac12\,s + m_\pi^2 \right)
= \frac{1}{\textcolor{black}{2}\,f_\pi^2}\left( -\frac12\,s + m_\pi^2 \right),    \nonumber 
\label{def-B-C-LO-s-wave}
\end{eqnarray}
and 
\begin{equation}
\mathcal{C}_{\LO}(s, t, u) = \mathcal{C}_{\LO}(t, u, s)  = \mathcal{C}_{\LO}(u, s, t) = 0\,. \nonumber
\end{equation}
\subsubsection{$S$-wave projection of scattering amplitude from NLO at tree-level diagrams}
At NLO, the $\mathcal{B}_{\NLO}$ and $\mathcal{C}_{\NLO}$ at $S$-wave level are read%
\footnote{In the following we suppress the integration boundaries of the $d(\cos\theta)$ integrals as they are always the same.}
\begin{eqnarray}
\mathcal{B}_{\NLO}^{(s)}(s) &\equiv& \textcolor{black}{\frac{1}{4}}\int d(\cos\theta)\,\mathcal{B}_{\NLO}(s, t, u) 
\nonumber\\
&=& \frac{1}{\textcolor{black}{2}\,f_\pi^4}\int d(\cos\theta)\left[\mathcal{B}_P(s, t, u) + \mathcal{B}_S(s, t-u) + \mathcal{B}_S(u, t-s) + \mathcal{B}_T(t)\right] ,
\nonumber\\
\mathcal{B}_{\NLO}^{(t)}(s) &\equiv& \textcolor{black}{\frac{1}{4}}\int d(\cos\theta)\,\mathcal{B}_{\NLO}(t,u,s) 
\nonumber\\
&=& \frac{1}{\textcolor{black}{4}\,f_\pi^4}\int d(\cos\theta)\left[\mathcal{B}_P(t,u,s) + \mathcal{B}_S(t,u-s) + \mathcal{B}_S(s,u-t) + \mathcal{B}_T(u)\right] ,
\nonumber\\
\mathcal{B}_{\NLO}^{(u)}(s) &\equiv& \textcolor{black}{\frac{1}{4}}\int d(\cos\theta)\,\mathcal{B}_{\NLO}(u,s,t) 
\nonumber\\
&=& \frac{1}{\textcolor{black}{4}\,f_\pi^4}\int d(\cos\theta)\left[\mathcal{B}_P(u,s,t) + \mathcal{B}_S(u,s-t) + \mathcal{B}_S(t, s-u) + \mathcal{B}_T(s)\right] ,
\nonumber
\end{eqnarray}
and
\begin{eqnarray}
\mathcal{C}_{\NLO}^{(s)}(s) &\equiv& \textcolor{black}{\frac{1}{4}}\int d(\cos\theta)\,\mathcal{C}_{\NLO}(s, t, u) 
\nonumber\\
&=& \frac{1}{\textcolor{black}{4}\,f_\pi^4}\int d(\cos\theta)\left[\mathcal{C}_P(s, t, u) + \mathcal{C}_S(s) + \mathcal{C}_T(t) + \mathcal{C}_T(u)\right]
\nonumber\\
\mathcal{C}_{\NLO}^{(t)}(s) &\equiv& \textcolor{black}{\frac{1}{4}}\int d(\cos\theta)\,\mathcal{C}_{\NLO}(t,u,s) 
\nonumber\\
&=& \frac{1}{\textcolor{black}{4}\,f_\pi^4}\int d(\cos\theta)\left[\mathcal{C}_P(t,u,s) + \mathcal{C}_S(t) + \mathcal{C}_T(u) + \mathcal{C}_T(s)\right]
\nonumber\\
\mathcal{C}_{\NLO}^{(u)}(s) &\equiv& \textcolor{black}{\frac{1}{4}}\int d(\cos\theta)\,\mathcal{C}_{\NLO}(u,s,t) 
\nonumber\\
&=& \frac{1}{\textcolor{black}{4}\,f_\pi^4}\int d(\cos\theta)\left[\mathcal{C}_P(u,s,t) + \mathcal{C}_S(u) + \mathcal{C}_T(s) + \mathcal{C}_T(t)\right]\,.\nonumber
\label{def-B-C-NLO-s-wave}
\end{eqnarray}
All relevant $S$-wave projections of $\mathcal{B}_P$ and $\mathcal{C}_P$  are given by
\allowdisplaybreaks
\begin{eqnarray}
\begin{aligned}
\mathcal{B}_P^{(s)}(s) & \equiv \textcolor{black}{\frac{1}{4}}\int d(\cos\theta)\,\mathcal{B}_P(s,t,u)
\nonumber \\
&= \textcolor{black}{\frac{1}{4}}\int d(\cos\theta)\left[\alpha_1\,m_\pi^4 +\alpha_2\,m_\pi^2\,t +\alpha_3\, t^2 +\alpha_4\,(s-u)^2 \right]
\nonumber\\
&= \textcolor{black}{\frac{1}{4}}\left[2\,\alpha_1\,m_\pi^4 -\,\alpha_2\,m_\pi^2\,s\,\sigma^2 +\frac{2}{3}\alpha_3\, s^2\,\sigma^4 +\alpha_4\left(2\,s^2 + 2\,s^2\,\sigma^2 +\frac23\,s^2\,\sigma^4 \right) \right]\,,
\nonumber\\
\mathcal{B}_P^{(t)}(s) &\equiv  \textcolor{black}{\frac{1}{4}}\int d(\cos\theta)\,\mathcal{B}_P(t, u, s)
\nonumber\\
&= \textcolor{black}{\frac{1}{4}}\int d(\cos\theta)\left[\alpha_1\,m_\pi^4 +\alpha_2\,m_\pi^2\,u +\alpha_3 u^2 +\alpha_4(t-s)^2 \right]
\nonumber\\
&= \textcolor{black}{\frac{1}{4}}\left[2\,\alpha_1\,m_\pi^4 -\,\alpha_2\,m_\pi^2\,s\,\sigma^2 +\frac{2}{3}\alpha_3\, s^2\,\sigma^4 +\alpha_4\left(2\,s^2 + 2\,s^2\,\sigma^2 +\frac23\,s^2\,\sigma^4 \right) \right],
\nonumber\\
\mathcal{B}_P^{(u)}(s) &\equiv  \textcolor{black}{\frac{1}{4}}\int d(\cos\theta)\,\mathcal{B}_P(u, s, t)
\nonumber\\
&= \textcolor{black}{\frac{1}{4}}\int d(\cos\theta)\left[\alpha_1\,m_\pi^4 +\alpha_2\,m_\pi^2\,s +\alpha_3\, s^2 +\alpha_4\,(u-t)^2 \right]
\nonumber\\
&= \textcolor{black}{\frac{1}{4}}\left[2\,\alpha_1\,m_\pi^4 + 2\,\alpha_2\,m_\pi^2\,s + 2\,\alpha_3\, s^2 + \frac23\,\alpha_4\,s^2\,\sigma^4 \right] ,
\nonumber
\end{aligned}
\end{eqnarray}
and
\begin{eqnarray}
\begin{aligned}
\mathcal{C}_P^{(s)}(s) &\equiv \textcolor{black}{\frac{1}{4}}\int d(\cos\theta)\,\mathcal{C}_P(s, t, u)
\nonumber\\
&= \textcolor{black}{\frac{1}{4}}\int d(\cos\theta)\left[\beta_1\,m_\pi^4 +\beta_2\,m_\pi^2\, s +\beta_3\, s^2 +\beta_4\,(t-u)^2 \right]
\nonumber\nonumber\\
&= \textcolor{black}{\frac{1}{4}}\left[2\,\beta_1\,m_\pi^4 + 2\,\beta_2\,m_\pi^2\, s + 2\,\beta_3\, s^2 + \frac23\,\beta_4\,s^2\,\sigma^4\right] \,,
\nonumber\\
\mathcal{C}_P^{(t)}(s) &\equiv  \textcolor{black}{\frac{1}{4}}\int d(\cos\theta)\,\mathcal{C}_P(t, u, s) 
\nonumber\\
&= \textcolor{black}{\frac{1}{4}}\int d(\cos\theta) \left[\beta_1\,m_\pi^4 +\beta_2\,m_\pi^2\, t +\beta_3\, t^2+\beta_4\,(u - s)^2 \right]
\nonumber\\
&= \textcolor{black}{\frac{1}{4}}\left[2\,\beta_1\,m_\pi^4 -\,\beta_2\,m_\pi^2\, s\,\sigma^2 +\frac{2}{3}\beta_3\, s^2\,\sigma^4+\beta_4
\left(2\,s^2 + 2\,s^2\,\sigma^2 +\frac23\,s^2\,\sigma^4 \right)\right] ,
\nonumber\\
\mathcal{C}_P^{(u)}(s) &\equiv  \textcolor{black}{\frac{1}{4}}\int d(\cos\theta)\,\mathcal{C}_P(u, s, t)
\nonumber\\
&= \textcolor{black}{\frac{1}{4}}\int d(\cos\theta)\left[\beta_1\,m_\pi^4 +\beta_2\,m_\pi^2\, u +\beta_3\, u^2
+\beta_4(s-t)^2 \right]
\nonumber\\
&= \textcolor{black}{\frac{1}{4}}\left[2\,\beta_1\,m_\pi^4 - \,\beta_2\,m_\pi^2\, s\,\sigma^2 +\frac{2}{3}\beta_3\, s^2\,\sigma^4+\beta_4
\left(2\,s^2 + 2\,s^2\,\sigma^2 +\frac23\,s^2\,\sigma^4 \right) \right]\,.\nonumber
\end{aligned}
\end{eqnarray}

\subsubsection{$S$-wave projection of scattering amplitude at NLO from one-loop diagrams}

Using the expressions calculated from Eq.~\eqref{eq:Jintegral}, we now can express the S-wave projection of the one-loop terms $\mathcal{B}_S$, $\mathcal{B}_T$, $\mathcal{C}_S$, and $\mathcal{C}_T$ as follows: 
\begin{eqnarray} 
\mathcal{B}_S^{(stu)}(s) &\equiv & \textcolor{black}{\frac{1}{4}}\int d(\cos\theta)\,\mathcal{B}_S(s,t-u)
\nonumber\\
&=& \textcolor{black}{\frac{1}{4}}\int d(\cos\theta)\,\bar{J}(s)\left[-\frac{1}{4}\,m_\pi^4 -\frac{1}{4}\,m_\pi^2\, s + \frac{1}{8}\,s^2 +\frac{1}{12}\left(m_\pi^2-\frac{s}{4}\right)(t-u)\right]
\nonumber\\
&=&\textcolor{black}{\frac{1}{2}}\bar{J}(s)\left[-\frac{1}{4}\,m_\pi^4 -\frac{1}{4}\,m_\pi^2\, s + \frac{1}{8}\,s^2 \right],
\nonumber\\
\mathcal{B}_S^{(uts)}(s) &\equiv & \textcolor{black}{\frac{1}{4}}\int d(\cos\theta)\,\mathcal{B}_S(u,t-s)
\nonumber\\
&=& \textcolor{black}{\frac{1}{4}}\int d(\cos\theta)\,\bar{J}(u)\left[-\frac{1}{4}\,m_\pi^4 -\frac{1}{4}\,m_\pi^2\, u + \frac{1}{8}\,u^2 +\frac{1}{12}\left(m_\pi^2-\frac{u}{4}\right)(t-s)\right]
\nonumber\\
&=& \textcolor{black}{\frac{1}{4}}\left[\frac{1}{6}\left( \frac{m_\pi^4}{2} - m_\pi^2\,s\right)\mathcal{J}_0(s) + \frac{1}{12}\left( \frac{s}{2} - 5\,m_\pi^2\right)\mathcal{J}_1(s) +\frac{7}{48}\,\mathcal{J}_2(s)\right]\,,
\nonumber\\
\mathcal{B}_S^{(tus)}(s) &\equiv & \textcolor{black}{\frac{1}{4}}\int d(\cos\theta)\,\mathcal{B}_S(t,u-s)
\nonumber\\
&=& \textcolor{black}{\frac{1}{4}}\int d(\cos\theta)\,\bar{J}(t)\left[-\frac{1}{4}\,m_\pi^4 -\frac{1}{4}\,m_\pi^2\,t + \frac{1}{8}\,t^2 +\frac{1}{12}\left(m_\pi^2-\frac{t}{4}\right)(u-s)\right]
\nonumber\\
&=& \textcolor{black}{\frac{1}{4}}\left[ \frac{1}{6}\left( \frac{m_\pi^4}{2} - m_\pi^2\,s\right)\mathcal{J}_0(s) + \frac{1}{12}\left( \frac{s}{2} - 5\,m_\pi^2\right)\mathcal{J}_1(s) +\frac{7}{48}\,\mathcal{J}_2(s)\right]\,,
\nonumber\\
\mathcal{B}_S^{(sut)}(s) &\equiv & \textcolor{black}{\frac{1}{4}}\int d(\cos\theta)\,\mathcal{B}_S(u,t-s)
\nonumber\\
&=& \textcolor{black}{\frac{1}{4}}\int d(\cos\theta)\,\bar{J}(s)\left[-\frac{1}{4}\,m_\pi^4 -\frac{1}{4}\,m_\pi^2\,s + \frac{1}{8}\,s^2 +\frac{1}{12}\left(m_\pi^2-\frac{s}{4}\right)(u-t)\right]
\nonumber\\
&=& \textcolor{black}{\frac{1}{2}}\bar{J}(s)\left[-\frac{1}{4}\,m_\pi^4 -\frac{1}{4}\,m_\pi^2\, s + \frac{1}{8}\,s^2 \right],
\nonumber\\
\mathcal{B}_S^{(ust)}(s) &\equiv & \textcolor{black}{\frac{1}{4}}\int d(\cos\theta)\,\mathcal{B}_S(u,s-t)
\nonumber\\
&=& \textcolor{black}{\frac{1}{4}}\int d(\cos\theta)\,\bar{J}(u)\left[-\frac{1}{4}\,m_\pi^4 -\frac{1}{4}\,m_\pi^2\,u + \frac{1}{8}\,u^2 +\frac{1}{12}\left(m_\pi^2-\frac{u}{4}\right)(s-t)\right]
\nonumber\\
&=& \textcolor{black}{\frac{1}{4}}\left[\frac{1}{6}\left( m_\pi^2\,s - \frac{7}{2}\,m_\pi^4 \right)\mathcal{J}_0(s) - \frac{1}{12}\left( \frac{s}{2} + m_\pi^2\right)\mathcal{J}_1(s) +\frac{5}{48}\,\mathcal{J}_2(s)\right]\,,
\nonumber\\
\mathcal{B}_S^{(tsu)}(s) &\equiv & \textcolor{black}{\frac{1}{4}}\int d(\cos\theta)\,\mathcal{B}_S(t,s-u)
\nonumber\\
&=& \textcolor{black}{\frac{1}{4}}\int d(\cos\theta)\,\bar{J}(t)\left[-\frac{1}{4}\,m_\pi^4 -\frac{1}{4}\,m_\pi^2\,t + \frac{1}{8}\,t^2 +\frac{1}{12}\left(m_\pi^2-\frac{t}{4}\right)(s-u)\right]
\nonumber\\
&=& \textcolor{black}{\frac{1}{4}}\left[\frac{1}{6}\left( m_\pi^2\,s - \frac{7}{2}\,m_\pi^4 \right)\mathcal{J}_0(s) - \frac{1}{12}\left( \frac{s}{2} + m_\pi^2\right)\mathcal{J}_1(s) +\frac{5}{48}\,\mathcal{J}_2(s) \right]\,,
\nonumber\\
\mathcal{B}_T^{(t)}(s) &\equiv  & \textcolor{black}{\frac{1}{4}}\int d(\cos\theta)\,\mathcal{B}_T(t)
\nonumber\\
&=&\textcolor{black}{\frac{1}{4}}\left[ -\frac{1}{8}\int d(\cos\theta)\,\bar{J}(t)\left[t-2\,m_\pi^2 \right]^2\right]
\nonumber\\
&=& \textcolor{black}{\frac{1}{4}}\left[-\frac{1}{2}\,m_\pi^4\,\mathcal{J}_0(s) + \frac{1}{2}\,m_\pi^2\,\mathcal{J}_1(s) -\frac{1}{8}\,\mathcal{J}_2(s) \right]\,,
\nonumber\\
\mathcal{B}_T^{(u)}(s) &\equiv & \textcolor{black}{\frac{1}{4}}\int d(\cos\theta)\,\mathcal{B}_T(u)
\nonumber\\
&=& \textcolor{black}{\frac{1}{4}}\left[-\frac{1}{8}\int d(\cos\theta)\,\bar{J}(u)\left[u-2\,m_\pi^2 \right]^2\right]
\nonumber\\
&=& \textcolor{black}{\frac{1}{4}}\left[-\frac{1}{2}\,m_\pi^4\,\mathcal{J}_0(s) + \frac{1}{2}\,m_\pi^2\,\mathcal{J}_1(s) -\frac{1}{8}\,\mathcal{J}_2(s)\right]\,,
\nonumber\\
\mathcal{B}_T^{(s)}(s) &\equiv & \textcolor{black}{\frac{1}{4}}\int d(\cos\theta)\,\mathcal{B}_T^{(s)}(s)
\nonumber\\
&=&\textcolor{black}{\frac{1}{4}}\left[-\frac{1}{8}\int d(\cos\theta)\, \bar{J}(s)\,(s-2\,m_\pi^2)^2\right]
\nonumber\\
&=&\textcolor{black}{\frac{1}{4}}\left[-\frac{1}{4}\,\bar{J}(s)\,(s-2\,m_\pi^2)^2\right]\,,
\nonumber  
\end{eqnarray}
and
\begin{eqnarray}
\mathcal{C}_S^{(s)}(s) &\equiv & \textcolor{black}{\frac{1}{4}}\int d(\cos\theta)\,\mathcal{C}_S(s) 
\nonumber\\
&=& \textcolor{black}{\frac{1}{4}}\int d(\cos\theta)\,\bar{J}(s)\left(\frac{1}{8}\,m_\pi^4 +\frac{1}{8}\, s^2\right)
\nonumber\\
&=& \textcolor{black}{\frac{1}{4}}\left[\frac14\,\bar{J}(s)\left(m_\pi^4 + s^2\right)\right],
\nonumber\\
\mathcal{C}_S^{(t)}(s) &\equiv  & \textcolor{black}{\frac{1}{4}}\int d(\cos\theta)\,\mathcal{C}_S(t)
\nonumber\\
&=& \textcolor{black}{\frac{1}{4}}\int d(\cos\theta)\,\bar{J}(t)\left(\frac{1}{8}\,m_\pi^4 +\frac{1}{8}\, t^2\right)
\nonumber\\
&=& \textcolor{black}{\frac{1}{4}}\left[\frac{1}{8}\,m_\pi^4\,\mathcal{J}_0(s) +  \frac{1}{8}\,\mathcal{J}_2(s)\right]\,,
\nonumber\\
\mathcal{C}_S^{(u)}(s) &\equiv & \textcolor{black}{\frac{1}{4}}\int d(\cos\theta)\,\mathcal{C}_S(u)
\nonumber\\
&=& \textcolor{black}{\frac{1}{4}}\int d(\cos\theta)\,\bar{J}(u)\left(\frac{1}{8}\,m_\pi^4 +\frac{1}{8}\, u^2\right)
\nonumber\\
&=& \textcolor{black}{\frac{1}{4}}\left[\frac{1}{8}\,m_\pi^4\,\mathcal{J}_0(s) +  \frac{1}{8}\,\mathcal{J}_2(s)\right]\,,
\nonumber\\
\mathcal{C}_T^{(t)}(s) &\equiv & \textcolor{black}{\frac{1}{4}}\int d(\cos\theta)\,\mathcal{C}_T(t) = \textcolor{black}{\frac{1}{4}}\left[ -\int d(\cos\theta)\,\mathcal{B}_T(t)\right]
\nonumber\\
&=& \textcolor{black}{\frac{1}{4}}\left[\frac{1}{2}\,m_\pi^4\,\mathcal{J}_0(s) - \frac{1}{2}\,m_\pi^2\,\mathcal{J}_1(s) + \frac{1}{8}\,\mathcal{J}_2(s)\right]\,,
\nonumber\\
\mathcal{C}_T^{(u)}(s) &\equiv & \textcolor{black}{\frac{1}{4}}\int d(\cos\theta)\,\mathcal{C}_T(u) = \textcolor{black}{\frac{1}{4}}\left[-\int d(\cos\theta)\,\mathcal{B}_T(u)\right]
\nonumber\\
&=& \textcolor{black}{\frac{1}{4}}\left[\frac{1}{2}\,m_\pi^4\,\mathcal{J}_0(s) - \frac{1}{2}\,m_\pi^2\,\mathcal{J}_1(s) + \frac{1}{8}\,\mathcal{J}_2(s)\right]\,,
\nonumber\\
\mathcal{C}_T^{(s)}(s) &\equiv & \textcolor{black}{\frac{1}{4}}\int d(\cos\theta)\,\mathcal{C}_T(s) = \textcolor{black}{\frac{1}{4}}\left[ -\int d(\cos\theta)\,\mathcal{B}_T(s) \right]
\nonumber\\
&=& \textcolor{black}{\frac{1}{4}}\left[\frac{1}{4}\,\bar{J}(s)(s-2\,m_\pi^2)^2\right]\,.\nonumber
\end{eqnarray}

\section{Discussion of other pole-finding methods}
\label{App:methods}

In this appendix, we comment on alternative pole-finding methods with various unitarization approaches.  After a brief explanation of the methods, we explain the reasons why they cannot be properly applied to our case.
While there are also methods that operate beyond the weak-binding assumption, such as the modern bootstrap method, a systematic investigation along those lines is beyond the scope of this work.

\subsubsection{The inverse amplitude method}

The inverse amplitude method (IAM) relies on an expansion of the amplitude as 
\begin{equation}
{1 \over T} =  {1 \over T_{\LO} + T_{\NLO}+\dots} \simeq  {1 \over T_{\LO}} - {T_{\NLO} \over T_{\LO}^2} + ... \, ,  
\end{equation}
which is supposed to be faithful for $|T_{\NLO}|\ll |T_{\LO}|$ (and a continuing hierarchy at higher orders), where $T_{\LO}$ and $T_{\NLO}$ are the respective LO and NLO contribution to the T-matrix amplitude.

The inverse of the exact amplitude $T^{-1} = \operatorname{Re}{T^{-1} + i \operatorname{Im}G}$, and the LO contribution $T_{\LO}$ is real.  As illustrated in Ref.~\cite{Oller:1998hw} (it defines the T-matrix operator via  ${\hat {\mathcal S}} \equiv 1- i{\hat {\mathcal T}} $, with an opposite sign to ours), now the amplitude can be written as 
\begin{eqnarray}
    T & = &  \frac{1}{ \operatorname{Re} T^{-1} + i \operatorname{Im} G } =  T_{\LO} \frac{1}{T_{\LO} \, (\operatorname{Re}  T^{-1} + i \operatorname{Im} G )\, T_{\LO} } T_{\LO} \notag\\
  &  \simeq &  T_{\LO} \, \frac{1}{T_{\LO}({1 \over T_{\LO}} - \operatorname{Re} {T_{\NLO}  \over T_{\LO}^2}   + i \operatorname{Im} G )T_{\LO} } \, T_{\LO}  \notag\\
    &  \simeq &  T_{\LO}\,  \frac{1}{T_{\LO} - \operatorname{Re}  T_{\NLO}    + i T_{\LO} \, \operatorname{Im} G \,T_{\LO} } \, T_{\LO}  \notag \\
 &  \simeq &  T_{\LO} \,\frac{1}{T_{\LO} -  T_{\NLO}  } \,T_{\LO}  \,,\label{eq:IAM}
 \end{eqnarray}
where for the last equality the approximation $\operatorname{Im} T_{\NLO} \simeq  - T_{\LO} \,\operatorname{Im} G \, T_{\LO}$ has been used. 
The IAM is built on the assumption that $T_{\LO} -  T_{\NLO}$ never gets close to zero. Bound states, however, are  defined by zeros of $T^{-1}$ implying the condition $T_{\LO} =  T_{\NLO} $ in~\eqref{eq:IAM}, where the series for $T^{-1}$ is no longer convergent and higher orders are parametrically as important as the terms kept in the IAM. In other words, the IAM is tailored to unitarize the amplitude in the physical region and describe resonances above threshold~\cite{Oller:1998hw, Oller:2024lrk}, but it is intrinsically ill–suited for a controlled search for subthreshold poles associated with bound states.

We may also note that, if $T_{\NLO} = - T_{\LO}  \,G \, T_{\LO}$, where Eq.~\eqref{eq:TNLO} in the main text yields $v_{\NLO} =0$,  Eq.~\eqref{eq:IAM} also goes back to the LO result as follows:  
\begin{equation}
    T  \simeq {v_{\LO} \over 1 + v_{\LO} \, G }  = {T_{\LO} \over 1 +  T_{\LO} \, G}\,. 
\end{equation}

\subsubsection{The N/D method}

There is also the generic N/D method, which is based on  separating the unitarity cut at $s>s_\text{th}=4m_\pi^2$ (in the denominator) and the unphysical cut at $s< s_{L}=0$   (in the numerator) by defining the T-matrix amplitude in the form 
\begin{equation}
    T(s) = \frac{N(s)}{D(s)}\,.
\end{equation}
For $s>s_\text{th}$, the numerator and denominator are related via 
$$ \operatorname{Im} D(s) = -{\sigma(s) \over 16\pi } N(s) + \sum_i \lambda (s_i) \delta(s-s_i)\, .$$
 The last term compensates the discontinuities when  $T(s)$ vanishes at $s=s_i$. A detailed derivation of this method can be found in \cite{Oller:2024lrk}; here we only provide a brief overview of the key results relevant for us.   

Taking advantage of the Cauchy integration theorem, for an arbitrary $s_0$ and an integer $n$, one may express $D(s)$ as
\begin{equation}
    D(s) = \frac{(s-s_0)^n}{\pi} \int^\infty_{s_{\rm th}}ds'  \frac{\operatorname{Im} D(s')}{(s'-s)(s'-s_0)^n} - \frac{(s-s_0)^n}{(n-1)!}\frac{d^{n-1}}{dz^{n-1}} \frac{D(z)}{z-s}|_{z= s_0} \,. \label{eq:Dsgeneral}
\end{equation}
In order to obtain a finite result from the integral on the right hand side of the equation, the asymptotic scaling needs to be $D(s)/s^n \to 0$ at $s\to \infty$. %
The choice of $n=1$ has been generally adopted for renormalizable theories in the literature.  One example is a study of the Higgs self-interaction~\cite{Cahn:1983vi},  adopting  $s_0 = -1$ as a preferred value.

In our effective field theory approach,  $n\ge 3$ is  needed in order to properly take into account the NLO contribution.  %
Using the fact that the last term of Eq.~\eqref{eq:Dsgeneral} is a polynomial of degree $(n-1)$ and choosing $n=3$, we reach 
\begin{equation}
    D_3(s)= -\frac{(s-s_0)^3}{16\pi^2} \int^\infty_{s_{\rm th}}ds'  \frac{\sigma(s)N(s) }{(s'-s)(s'-s_0)^3}+  a_0 + a_1 (s-s_0) + a_2 (s-s_0)^2  
    + \sum_i {\gamma_i \over s - s_i} \,. \label{eq:Ds3}
\end{equation}
Now the high-energy behavior of the integral can be regularized by the denominator of the integrating factor. Moreover, the value of $a_0$ may be fixed by setting a normalization condition $D(s)|_{s = s_0} =1$ after neglecting the discontinuities. Nevertheless, there is no further input to find out the values of $a_1$ and $a_2$.  We have checked that the uncertainties in $a_1$ and $a_2$ eliminate the predictive power of this method.

\bibliography{DM-B-ref}

\end{document}